\documentclass[12pt,floatfix]{iopart}
\usepackage{graphicx}
\usepackage{iopams}
\usepackage{cite}

\begin{document}
\jl{2}
\title{Structure and properties of small sodium clusters.}
\author{Ilia A Solov'yov\footnote[1]{E-mail:
solovyov@rpro.ioffe.rssi.ru; on leave from St.Petersburg State Technical
University, Politechnicheskaya 29, 195251 St.Petersburg, Russia},
Andrey V Solov'yov\footnote[2]{E-mail:
solovyov@th.physik.uni-frankfurt.de; on leave from A.F.Ioffe Physical-Technical Institute,
Russian Academy
of Sciences, Politechnicheskaya 26, St. Petersburg 194021,
Russia}
and
Walter Greiner\footnote[3]{E-mail:
greiner@th.physik.uni-frankfurt.de}
}

\address{Institut f\"ur Theoretische Physik der Johann Wolfgang
Goethe-Universit\"at, 60054 Frankfurt am Main, Germany}

\begin{abstract}
We have investigated
structure and properties of small metal clusters using
all-electron {\it ab initio} theoretical methods based on
the Hartree-Fock approximation and density functional theory,
perturbation theory and compared
results of our calculations with the available experimental
data and the results of other theoretical works.
We have systematically calculated the optimized geometries of neutral
and singly charged sodium clusters having up to 20 atoms, their
multipole moments (dipole and quadrupole), static polarizabilities,
binding energies per atom, ionization potentials and frequencies
of normal vibration modes. 
Our calculations demonstrate the
great role of many-electron correlations in the formation of
electronic and ionic structure of small metal clusters and form a good basis
for further detailed study of their dynamic properties, as
well as structure and properties of other atomic cluster systems.
\end{abstract}

\section{Introduction}

Atomic clusters and small nanoparticles have been recognized as new
physical objects with their own properties relatively
recently. This became clear after such experimental successes as
the discovery of electron shell structure in metal clusters
\cite{Knight84}, observation of plasmon resonances
in metal clusters \cite{Br89,Selby89,Selby91} and fullerenes
\cite{Bertsch92,Hertel92},
formation of singly and doubly charged negative cluster ions
\cite{Lutz} and many more others.
The novelty of cluster physics is also greatly connected with
the fact that cluster properties explain
the transition from
single atoms or molecules to solid state.
Comprehensive survey of the field can be found 
in review papers and books, see e.g.
\cite{deHeer93,Brack93,BrCo94,Haberland94,Guet97,MetCl99,LesHouches}. 
              
There are many different types of clusters, such as
metallic clusters, fullerenes, molecular clusters,
semiconductor clusters, organic clusters,
quantum dots, positively and negatively charged clusters, which
all have their own features and properties.
In this paper we focus on the detailed systematic study of the structure
and properties of small metal clusters and in particular
sodium clusters using {\it ab initio} all-electron many-body
theory methods.  

So far, for sodium clusters, systematic calculations of
cluster properties  on the same level
of theory as in our present work (i.e. all electron {\it ab initio}) 
have been performed only for clusters with $N \leq 10$ 
\cite{MetCl99,Boustani87,Bonacic88,Boustani88,Guest88,Rayane99},
where $N$ is a number of atoms in a cluster.
In our work we extend this limit up to $N \leq 20$. Note that
most of the cited papers are focused on the investigation of neutral
cluster properties rather than ions. In our present work we perform
systematic comparative analysis of properties of neutral and singly charged
sodium clusters in the specified size range.

During the last decade, there were performed numerous experimental
and theoretical investigations of the properties of small metal clusters
as well as the processes with their involvement. Here we are not able
to review even all essential results obtained in the field 
and only refer to those, which are related the most closely to
the subject of our paper. In \cite{Knight84}, it was experimentally
proved that metal clusters have the shell electronic structure and
the magic cluster numbers have been determined by observation 
of the sodium cluster abundances in mass spectra.
Experimental study of electronic structure
and properties of small metal clusters have been performed in
\cite{Akeby90,Knight85} (for review also see
\cite{deHeer93,BrCo94,Haberland94,MetCl99,LesHouches}).
In \cite{Akeby90}, there have been measured the ionisation potentials
for a sequence of small neutral and positively charged sodium metal clusters,
which independently proved their shell structure.
The dipole polarizabilities of sodium clusters have been experimentally
determined in \cite{Knight85}. Dissociation energies of neutral and positively
charged small sodium and potassium
metal clusters have been measured in \cite{Br89a,Br90,Br91}.
Dynamical properties of clusters have been studied by means of photon,
electron and ion scattering. These methods are the 
traditional tools for probing properties and internal
structure of various physical objects. Using these methods, for example,
plasmon excitations in metal clusters
\cite{Br89,deHeer87} and fullerenes \cite{Hertel92}
have been observed (for review also see \cite{deHeer93,BrCo94}).

Metal clusters have also been  studied  theoretically.
Structural properties of small metal clusters have
been widely investigated using quantum chemistry methods.
Here we refer to the papers 
\cite{Boustani87,Bonacic88,Boustani88,Guest88,Martins85,Spieg98,
Nagueira99,Gutierrez01},
in which optimized geometries, binding energies, ionization potentials, 
electron structure and electron transport properties
of small lithium and sodium clusters have been calculated.
In these papers the systematic analysis of the cluster properties has been 
limited by cluster sizes  $N \leq 10$.  In the present paper we extend this
limit  up to $N \leq 20$ and perform systematic analysis of
various cluster characteristics both for neutral clusters
and singly charged cluster ions.

In a last few years, a number of papers
have been devoted to the
calculation of  dipole static polarizabilities of neutral sodium and
lithium clusters \cite{Rayane99,Blundell00,Kronik00,Kummel00,Manninen00}.
Note that most of these studies have been performed within the cluster 
size range
$N \leq 20$. The results of different theoretical approaches have
been compared with the experimental data  from \cite{Knight85}.
However, only in \cite{Rayane99}, calculations  of the 
cluster geometries and polarizabilities have been performed 
on the same level of theory (i.e. all electron {\it ab initio})  
as in our work and were limited by $N\leq 8$.

Alternatively, the jellium model for metal clusters was
suggested. This model explains well enough the shell
structure of metal clusters
and their essential dynamic properties, such as plasmon excitations.
Initially, jellium calculations for metal clusters
were based on the density functional formalism with the use of pseudopotentials
for the description of electron relaxation effects and lattice structure
\cite{Martins81}. Fully self-consistent calculations for spherical
jellium metal clusters
have been performed within the framework of the spin-density-functional method
\cite{Hintermann83} and the Kohn-Sham formalism
for the self-consistent determination of electron wave functions
\cite{Ekky84,Ekky85}. The Hartree-Fock scheme for the self-consistent
determination of the electron wave functions of spherical jellium
metal clusters was also introduced later in
\cite{GJ92,IIKZ93}. This approach was generalized for axially
deformed cluster systems in \cite{LSSCG00}.
Dynamical jellium model for
metal clusters, which treats simultaneously collective vibrational modes
(volume vibrations, i.e. breathing, plus shape vibrations) of the ionic jellium
background in a cluster, the quantized electron motion and
interaction between the electronic and ionic subsystems
was developed in \cite{GSG99,GISG00}.
                   
The jellium model provides a very useful basis for studying
various collision processes, such
as photabsorption \cite{Ivanov96}, photoionization
\cite{Bertsch92,Bertsch91,C20-C60},
elastic \cite{GCSG97,GEMS98}
and inelastic scattering \cite{GEMS98,GIS97,GISG98,GIPS00},
electron attachment \cite{CGIS98,CGIS99},
photon emission \cite{GS97,GIS98} and others,
involving metal clusters,.  On the basis of the
jellium model
one can develop {\it ab initio} many-body theories, such as the
random phase approximation with exchange  or the
Dyson equation method 
and effectively solve many-electron correlation problem
even for relatatively large cluster systems containing
up to 100 atoms or even more. Review of these methods in
their application to the electron scattering of metal clusters
one can find in \cite{AVSol}.
As elucidated in the papers cited above,
many-electron correlations are quite
essential for the correct description of various
characteristics of the cluster systems.

In spite of the fact that the jellium model with all its
modifications is rather successful in explaining numerous
phenomena involving metal clusters it obviously has its
limits, because this model does not take into account the
detailed ionic structure of clusters.
The correspondence between predictions of the
jellium model and the results of more advanced quantum
chemistry calculations have not been performed in a systematic
way so far. Partially, this is connected with the fact
that quantum chemistry calculations are usually limited
by small sizes of clusters, while the jellium
model becomes adequate for larger cluster systems.
Knowledge of the ranges of applicability of the jellium model
and the level of its accuracy is important, because
the jellium model often gives much more efficient theoretical
basis particularly, when dealing with larger cluster systems. 

In this paper we have undertaken  detailed systematic
theoretical study of structure and properties of sodium
clusters beyond the jellium model 
using all-electron {\it ab initio} theoretical methods
based on the Hartree-Fock approximation, density functional theory and
perturbation theory, for clusters that size is large enough for
jellium calculations.
Namely, we have calculated optimized  geometries of neutral
and singly-charged sodium clusters consisting of up to 20 atoms, their
multipole moments (dipole and quadrupole), static polarizabilities,
binding energies per atom, ionization potentials and frequencies
of normal vibration modes.   
We compare results of our calculations with the available experimental
data, results of other theoretical works performed both
within the framework of the jellium model and beyond, using  quantum chemistry
methods, and elucidate the level of accuracy of different
theoretical approaches.  Also, we 
demonstrate the great role of many-electron correlations in the
formation of structure and properties of small metal clusters.
Our results form a good basis
for the detailed study of dynamic properties of small metal clusters as well
as structure and properties of other atomic cluster systems.

Our calculations elucidate
the level of accuracy of various theoretical schemes for the treatment of
electronic structure in metal clusters, which is important to
know and is not obvious in advance due to complexity of theoretical
methods involved.
Some characteristics (dipole and quadrupole moments or spectra of
normal vibration modes, for example),
which we have calculated in this paper are new and were not studied
before, at least according to our knowledge.
These characteristics, however,
might be, important, for instance, when considering dynamics
of a cluster beam
in an external non-homogeneous electric or magnetic field.
Indeed, namely, cluster multipole moments  should be responsible
for the cluster isomers separation in the non-homogeneous external fields.
We analyse the connection between the principal values of the
cluster quadrupole moments tensor and
the cluster shape (oblate, prolate or triaxially deformed).

The frequencies of the surface and volume vibration modes have been
determined in the spectra of the cluster normal vibration frequencies
and their correspondence to the predictions of the dynamical jellium
model \cite{GSG99,GISG00} was established.

Our calculations have been performed with the use of 
the Gaussian 98 software package \cite{Gaussian98}.   
We have used the atomic system of units in this  paper,
$\hbar=m_e=|e|=1$ unless other units are not indicated.

\section{Theoretical methods}
\label{theory}

In this work we are studying  structure and properties of small
sodium clusters on the basis of all-electron {\it ab initio} many-body theory
methods. We calculate the optimized geometries of clusters
consisting of up to $N \leq 20$ atoms, where $N$
is the number of atoms in the cluster.
For the sequence of clusters with $N \leq 20$, we determine
size dependence of the cluster ionization potentials, total
energies, multipole moments (dipole and quadrupole), bonding distances and
dipole polarizabilities. We also calculated and analyze vibration spectra  
of the clusters.

We  have done these calculations using different theoretical
schemes.  We have calculated cluster characteristics
in the all-electron Hartree-Fock approximation. This approximation does not
take into account many-electron correlations in the system,
which turn out to play essential role in the formation of
clusters properties. Therefore,  we also calculate all the
characteristics using post Hartree-Fock theories accounting
for many-electron correlations. Namely, this was done in
the M\o ller and Plesset perturbation theory of the second
and the fourth order and the three parameter Becke's
gradient-corrected exchange functional with the gradient-corrected
correlation functional of Lee, Yang and Parr.

Below, we discuss theoretical methods used in our work. 
The aim of this discussion is to present essential ideas
of the methods and give the necessary references, rather than
to describe them in detail.

\subsection{Hartree-Fock method (HF)}

In the Hartree-Fock approximation,
the many-electron wave function of a cluster is expressed
as antisymmetrized product of the single-electron wave functions,
$\psi_i$, 
of cluster electrons, which are also often called
molecular orbitals.
The Hartree-Fock equation for the determination
of the  molecular orbitals $\psi_i$ reads as
(see e.g. \cite{Lindgren}):

\begin{equation}
\left( - \Delta/2 + U_{ions} + U_{HF}\right) \psi_i \, = \,
\varepsilon_i \psi_i.
\label{HF}
\end{equation}

\noindent
Here, the first term represents the kinetic energy of the $i$-th electron,
and $U_{ions}$ describes its attraction to the ions in the cluster.
The Hartree-Fock
potential $U_{HF}$ represents the Coulomb and the exchange interaction
of the electron $i$ with other electrons in the cluster, 
$\varepsilon_i$ is the single electron energy.

In {\it Gaussian 98}, 
the molecular orbitals, $\psi_i$, are approximated by a linear combination
of a pre-defined set
of single-electron functions, $\chi_{\mu}$,  known as basis functions.
This expansion reads as follows:

\begin{equation}
\psi_i=\sum_{\mu = 1}^{N} c_{\mu i} \chi_{\mu}, 
\label{phi_i}
\end{equation}

\noindent
where coefficients c$_{\mu i}$ are the molecular orbital expansion
coefficients, $N$ is the number of basis functions, which are chosen
to be normalized. 

The basis functions $\chi_{\mu}$ are defined as 
linear combinations of primitive gaussians: 

\begin{equation}
\chi_{\mu}=\sum_{p} d_{\mu p}g_p,
\label{hi_mu}
\end{equation}

\noindent
where $d_{\mu p}$ are  fixed constants within a given basis set,
the  primitive gaussians, $g_p= g(\alpha , {\bf r})$, are
the gaussian-type atomic functions having  the following form:

\begin{equation}
g(\alpha , {\bf r})=cx^ny^mz^l \e^{-\alpha r^2}
\label{g}
\end{equation}

\noindent
Here, $c$ is the normalization constant. The choice of
the integers $n$, $m$ and $l$ defines the type of the primitive
gaussian function: s, p, d or f (for details see \cite{Gaussian98_man}).

Substituting these expansions in the Hartree-Fock equations (\ref{HF}),
one can rewrite them in the form, known also
as the Roothaan and Hall equations:

\begin{equation}
\sum_{\nu =1}^N(H_{\mu\nu}-\varepsilon_iS_{\mu\nu})c_{\nu i}=0  \\
\mu=1,2,...,N
\label{Roothaan}
\end{equation}

\noindent
Being written in the matrix form, this equation reads as:

\begin{equation}
HC=SC\varepsilon,
\label{Roothaan_matrix}
\end{equation}

\noindent
where each element is a matrix. Here, $\varepsilon$ is a diagonal matrix
of orbital energies,  each of its elements $\varepsilon_i$ is the
single-electron energy of the molecular orbital $\psi_i$, H is
the Hamiltonian matrix as it follows from  (\ref{HF}), S is the overlap matrix,
describing the overlap between orbitals. For more details 
regarding this formalism see \cite{Gaussian98_man}.

Equations (\ref{Roothaan_matrix})
are none linear and must be solved iteratively.
The procedure which does so is called
the {\it Self-Consistent Field} (SCF) method.

The above written  equations  consider the restricted Hartree-Fock
method.  For the open shell systems, the unrestricted Hartree-Fock
method has to be used. In this case, the alpha and beta electrons
with spins up and down
are assigned to different orbitals, resulting in two sets of molecular
orbital expansion coefficients:

\begin{eqnarray}
\psi_i^{\alpha} =\sum_{\mu = 1}^{N} c_{\mu i}^{\alpha} \chi_{\mu}
\nonumber
\\
\psi_i^{\beta} =\sum_{\mu = 1}^{N} c_{\mu i}^{\beta} \chi_{\mu}, 
\label{phi_i_unrestr}
\end{eqnarray}

The two sets of coefficients result in two sets of the Hamiltonian matrices and
the two sets of orbitals.

\subsection{M\o ller-Plesset perturbation theory method ($MP_n$)}

The Hartree-Fock theory provides an inadequate treatment of 
electrons motion within a molecular system, because
it does not properly treat many-electron correlations.
The many electron correlations can be accounted for using
different methods. The most straightforward way for achieving this goal
is based on the perturbation theory. Indeed, the total Hamiltonian, $H$, of the
cluster can be divided into two parts

\begin{equation}
H= H_0 +  V
\label{MP_base}
\end{equation}

\noindent Here $ H_0$ is the Hamiltonian corresponding to the Hartree-Fock
level of theory and $V$ is the residual interelectron interaction, which
can be treated as a small perturbation.

Considering $V$ as a small perturbation one can construct the solution
of the Schr\"odinger equation for many-electron system in an arbitrary order
of the perturbation theory.
The perturbation theory of this type is well known
since the work by M\o ller-Plesset \cite{MP} and can be found in
numerous textbooks on quantum mechanics (see e.g. \cite{LL3}).

%
%

Below we refer to this theoretical method as to the 
M\o ller-Plesset perturbation theory \cite{MP} of the second or forth order,
$MP_2$ or $MP_4$. Indices here indicate the order of the perturbation
theory.

\subsection{Density functional methods (B3LYP)}

The density functional theory (DFT) is based upon a
strategy of modelling electron
correlation  via general functionals of the electron density.
Within the DFT one has to
solve the Kohn-Sham equations, which read as
(see e.g. \cite{deHeer93,Brack93,Haberland94,Guet97,MetCl99,LesHouches})

\begin{equation}
\left( \frac{\hat p^2}2+U_{ions}+V_{H}+V_{xc}\right)
\psi_i =\varepsilon _i \psi _i,
\end{equation}
where the first term represents the kinetic energy of the $i$-th electron,
and $U_{ions}$ describes its attraction to the ions in the cluster,
$V_{H}$ is the Hartree part of the interelectronic interaction:
\begin{equation}
V_{H}(\vec r)=\left. \int \frac{\rho(\vec r\,')}{|\vec r-\vec r\,'|}
\, d\vec r\,'\right.,
\end{equation}
and $\rho(\vec r\,')$ is the  electron density:
\begin{equation}
\rho(\vec r)=\sum_{\nu=1}^{N} \left|\psi_i(\vec r) \right|^2,
\end{equation}

\noindent
where $V_{xc}$ is the local exchange-correlation potential,
$\psi_i$ are the electronic orbitals and $N$ is the number of 
electrons in the cluster.

The exchange-correlation potential is defined as the functional
derivative of the exchange-correlation energy functional:
\begin{equation}
V_{xc}=\frac{\delta E_{xc}[\rho]}{\delta \rho(\vec r)},
\end{equation}

The approximate functionals
employed by DFT methods partition the exchange-correlation energy
into two parts, referred to as
{\it exchange} and {\it correlation} parts:

\begin{equation}
E_{xc}[\rho]= E_x(\rho)+E_c(\rho)
\label{ex_core}
\end{equation}

Physically, these two terms correspond to same-spin and
mixed-spin interactions, respectively. Both parts are the
functionals of the electron density, which can be of
two distinct types: either {\it local} functional  depending on only
the electron density $\rho$ or {\it gradient-corrected} functionals
depending on both $\rho$ and its gradient, ${\bf \nabla} \rho$.

In literature, there is a variety of exchange correlation functionals.
Below, we refer only to those, which are related to the calculation
performed in this work.

The local exchange functional
is virtually always defined as follows:

\begin{equation}
E^{LDA}_x=-\frac{3}{2} (\frac{3}{4\pi})^{1/3} \int \rho^{4/3}d^3 {\bf r}
\label{LDA}
\end{equation}

This form was developed to reproduce the exchange energy of a uniform
electron gas. By itself, however, it is not sufficient for
the adequate description of atomic clusters. 

The gradient-corrected exchange functional introduced by Becke \cite{Becke88}
and based on the LDA exchange functional reads as:

\begin{equation}
E^{B88}_x=
E^{LDA}_x-\gamma\int\frac{\rho^{4/3}x^2}{1+6\gamma sinh^{-1}x}d^3 {\bf r}
\label{becke}
\end{equation}

\noindent
where x=$\rho^{-4/3}|\nabla\rho|$ and $\gamma=0.0042$ is a parameter chosen
to fit the known exchange energies of the noble gas atoms.

Analogously to the above written exchange functionals,
there are local and gradient-corrected correlation functionals,
for example,  those introduced  by Perdew and Wang  \cite{PerWan} or
by Lee, Yang and Parr \cite{LYP}.
Their explicit expressions are somewhat lengthy and thus we 
do not present them here and  refer to the original papers.

In the pure DFT, an exchange functional usually pairs with a
correlation functional.
For example, the well-known BLYP functional pairs
Becke's gradient-corrected exchange functional (\ref{becke})
with the gradient-corrected
correlation functional of Lee, Yang and Parr \cite{LYP}.

In spite of the success of the pure DFT theory in many cases, one
has to admit that the Hartree-Fock theory accounts for
the electron exchange the most naturally and precisely.
Thus, Becke has suggested \cite{Becke88} functionals which include a mixture
of Hartree-Fock
and DFT exchange along with DFT correlations, conceptually defining $E_{xc}$
as:

\begin{equation}
E^{mix}_{xc}=c_{HF}E^{HF}_x + c_{DFT}E^{DFT}_{xc},
\label{becke_hyb}
\end{equation}

\noindent
where $c_{HF}$ and $c_{DFT}$ are  constants.
Following this idea, a 
Becke-type three parameter functional can be defined  as follows:

\begin{eqnarray}
E_{xc}^{B3LYP}&=&E^{LDA}_x + c_0(E^{HF}_x - E^{LDA}_X) +
c_x  (E^{B88}_x- E^{LDA}_x) +
\nonumber
\\
&+&E^{VWN3}_c + c_c(E^{LYP}_c - E^{VWN3}_c)
\label{b3lyp}
\end{eqnarray}

\noindent

Here, $c_0=0.2$, $c_x=0.72$ and  $c_c=0.81$ are constants, which
were defined by fitting to the atomization energies, 
ionization potentials, proton affinities and
first-row atomic energies \cite{Gaussian98_man}.
$E^{LDA}_x$ and $E^{B88}_x$ are defined in (\ref{LDA}) and
(\ref{becke}) respectively. $E^{HF}_x$ is the functional
corresponding to Hartree-Fock equations (\ref{HF}).
The explicit form for the correlation functional $E^{VWN3}_c$
as well as for 
gradient-corrected correlation functional of Lee, Yang and Parr,
$E^{LYP}_c$, one can find in \cite{VWN} and \cite{LYP} correspondingly.
Note that instead of $E^{VWN3}_c$ and $E^{LYP}_c$ in (\ref{b3lyp})
one can also use the Perdew and Wang correlation functional \cite{PerWan}.

\subsection{Geometry optimization}
\label{Geometry}

The cluster geometries, which we have calculated in our work,
have been determined using the geometry optimization procedure.
This procedure implies the calculation of the multidimensional
potential energy surface for a cluster and then finding
local minima on this surface. 
The key point for this search
is fixing the starting geometry of the cluster, which could 
converge during the calculation to the local or global minimum.
There is no unique way in achieving this goal with {\it Gaussian 98}.

In our
calculations, we have created the starting geometries empirically,
often assuming certain cluster symmetries. Note, that during
the optimization process the geometry of the cluster as well
as its initial symmetry sometimes change dramatically.
All the characteristics of clusters, which we have calculated
and presented in next section, are obtained for the
clusters  with optimized geometry.

In our calculations, we have
made no assumptions on the core electrons in the optimized clusters,
which means that all electrons available in the system, have
been taken into account, when computing potential energy surface.
For clusters with $N >10$, this process becomes rather computer
time demanding. Thus, in this work we have limited our calculations
by clusters consisting up to $N \leq 20$.

\subsection{Normal vibrations}

Knowledge of the potential energy surface in the vicinity
of a local  minimum, allows one easily to determine corresponding normal
vibration modes of the system. We have performed such calculation
and determined the vibration energy spectrum for a number of clusters.
Particular attention in this calculation has been paid to the
identification of the breathing and the surface vibration modes and
comparison their frequencies with those predicted in \cite{GSG99,GISG00}
for spherical sodium clusters on the basis of the dynamical jellium model.

\section{Results of calculations and discussion}

In this section we present the results of calculations performed
with the use of methods described above. We have calculated
the optimized  geometries of neutral
and singly charged sodium clusters consisting of up to 20 atoms, their
multipole moments (dipole and quadrupole), static polarizabilities,
binding energies per atom, ionization potentials and frequencies
of the normal vibration modes.   
We compare results of our calculations with the available experimental
data and the results of other theoretical works performed both
within the framework of the jellium model and beyond, using  quantum chemistry
methods and establish the level of accuracy of different
theoretical approaches. Particular attention is paid to the
clusters in the range  $10 < N < 20$, because some
characteristics  of the clusters in this size range
have been calculated on the {\it ab initio} basis in
our paper for the first time.
Also, we demonstrate the great role of many-electron correlations in the
formation of structure and properties of small metal clusters.

\subsection{Geometry optimization of Na$_{n}$ and Na$_{n}^{+}$ clusters}
\label{geom_opt}

\begin{figure}[h]
\begin{center}
\includegraphics[scale=0.6]{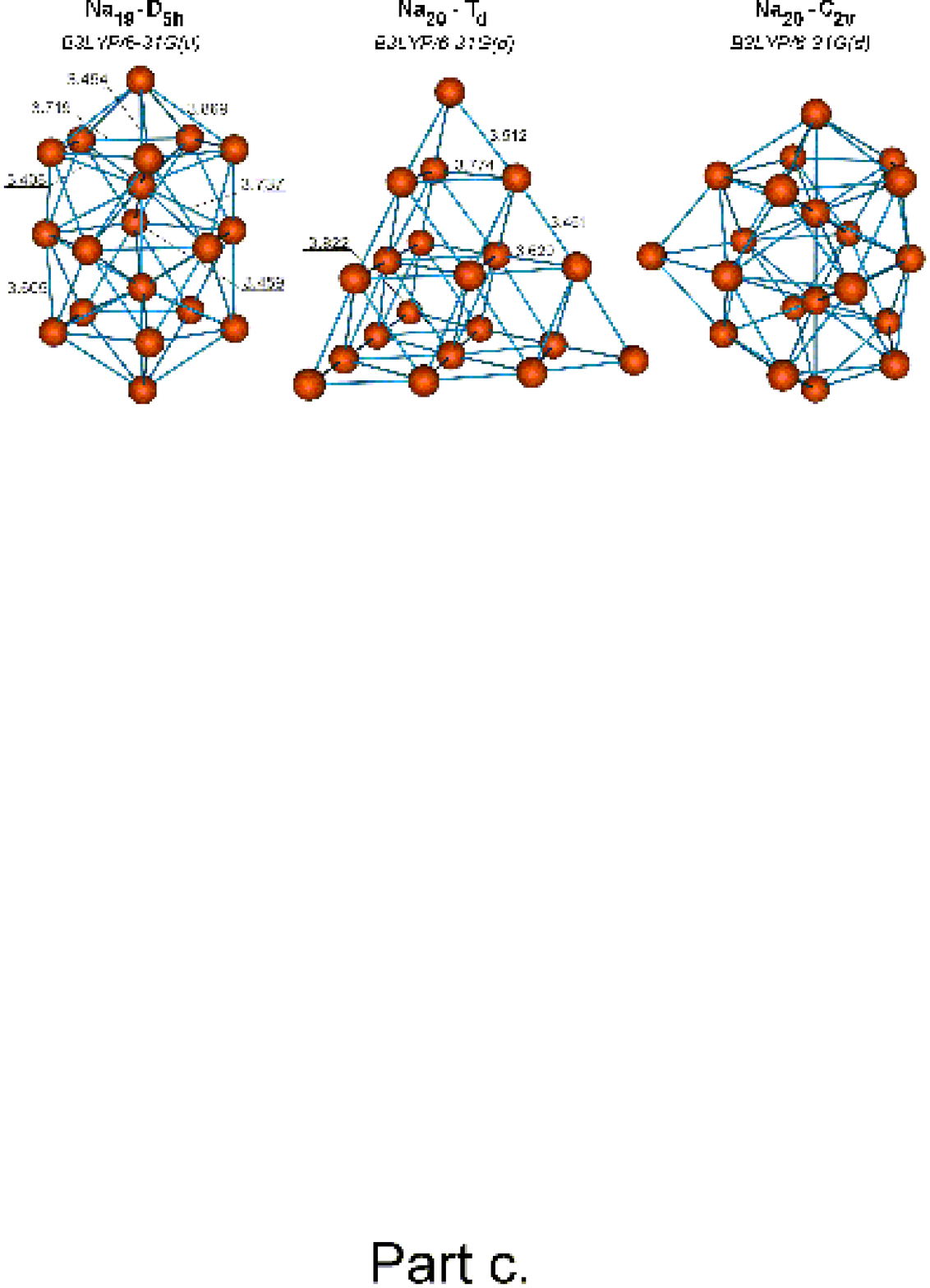}
\end{center}
\caption{
Optimized geometries of neutral sodium clusters
$Na_2 - Na_{10}$ (part a), $Na_{11}- Na_{18}$ (part b) 
and $Na_{19}- Na_{20}$ (part c).
The interatomic distances are given in angstroms.
The label above
each cluster image indicates its point symmetry group and
the calculation method by 
which the cluster was optimized.
}
\label{geom_neutral}
\end{figure}

\begin{figure}[h]
\begin{center}
\includegraphics[scale=0.6]{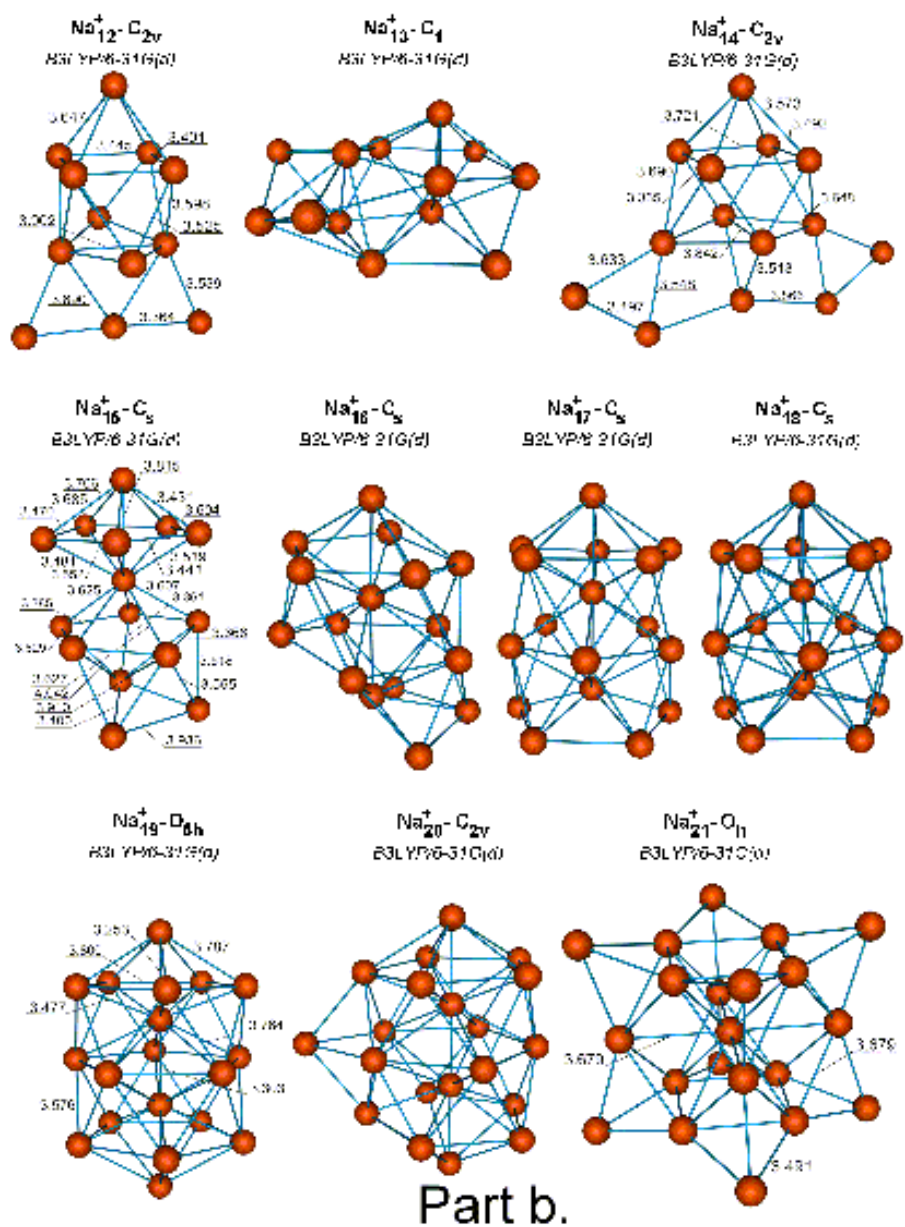}
\end{center}
\caption{
Optimized geometries of singly charged sodium clusters
$Na_2^+ - Na_{11}^+$ (part a) and $Na_{12}^+- Na_{21}^+$ (part b).
The interatomic distances are given in angstroms.
The label above
each cluster image indicates the point symmetry group and
the calculation method by 
which the cluster was optimized.}
\label{geom_ion}
\end{figure}

Results of the cluster geometry optimization for neutral and
singly charged  sodium clusters consisting of up to 20 atoms shown in
figures \ref{geom_neutral} and \ref{geom_ion}  respectively.
The cluster geometries have been
determined using the methodology described in section \ref{theory}.
Namely, the optimization of the cluster geometries
has been performed with the use of $B3LYP$ and $MP_2$ methods.

\newpage
\begin{figure}[h]
\begin{center}
\includegraphics[scale=0.6]{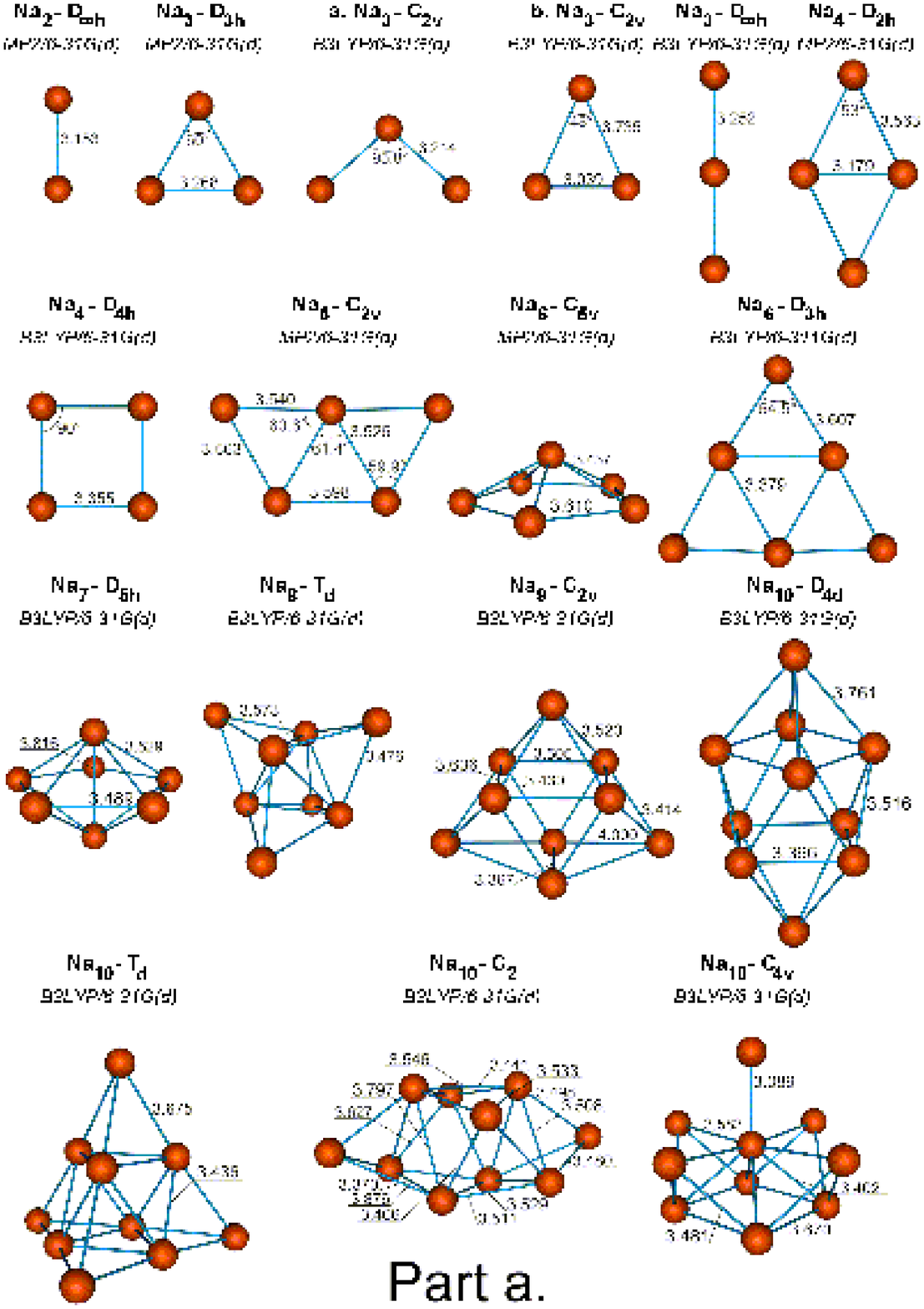}
\end{center}
\end{figure}

\newpage
\begin{figure}[h]
\begin{center}
\includegraphics[scale=0.6]{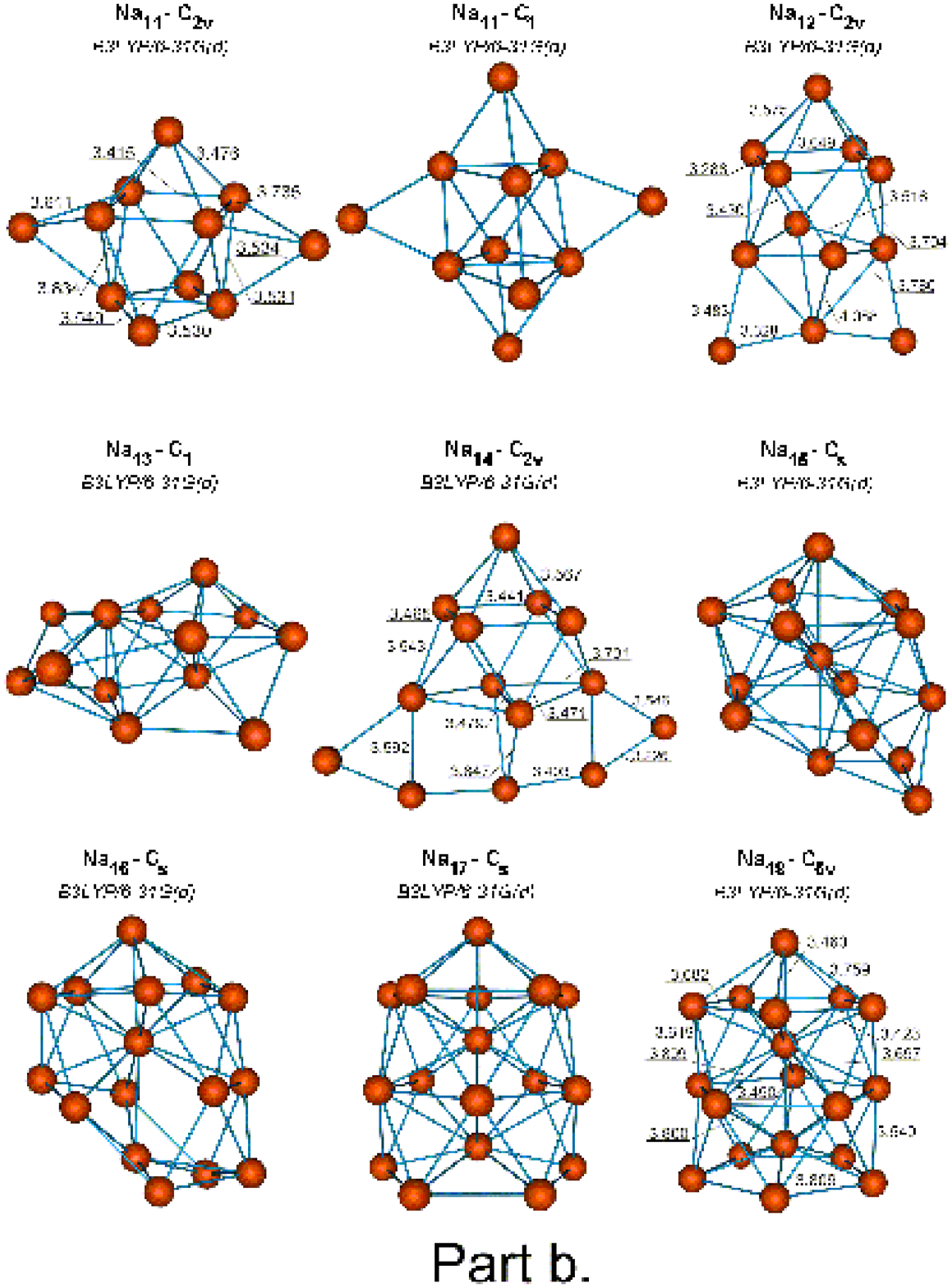}
\end{center}
\end{figure}

\newpage

\newpage
\begin{figure}[h]
\begin{center}
\includegraphics[scale=0.6]{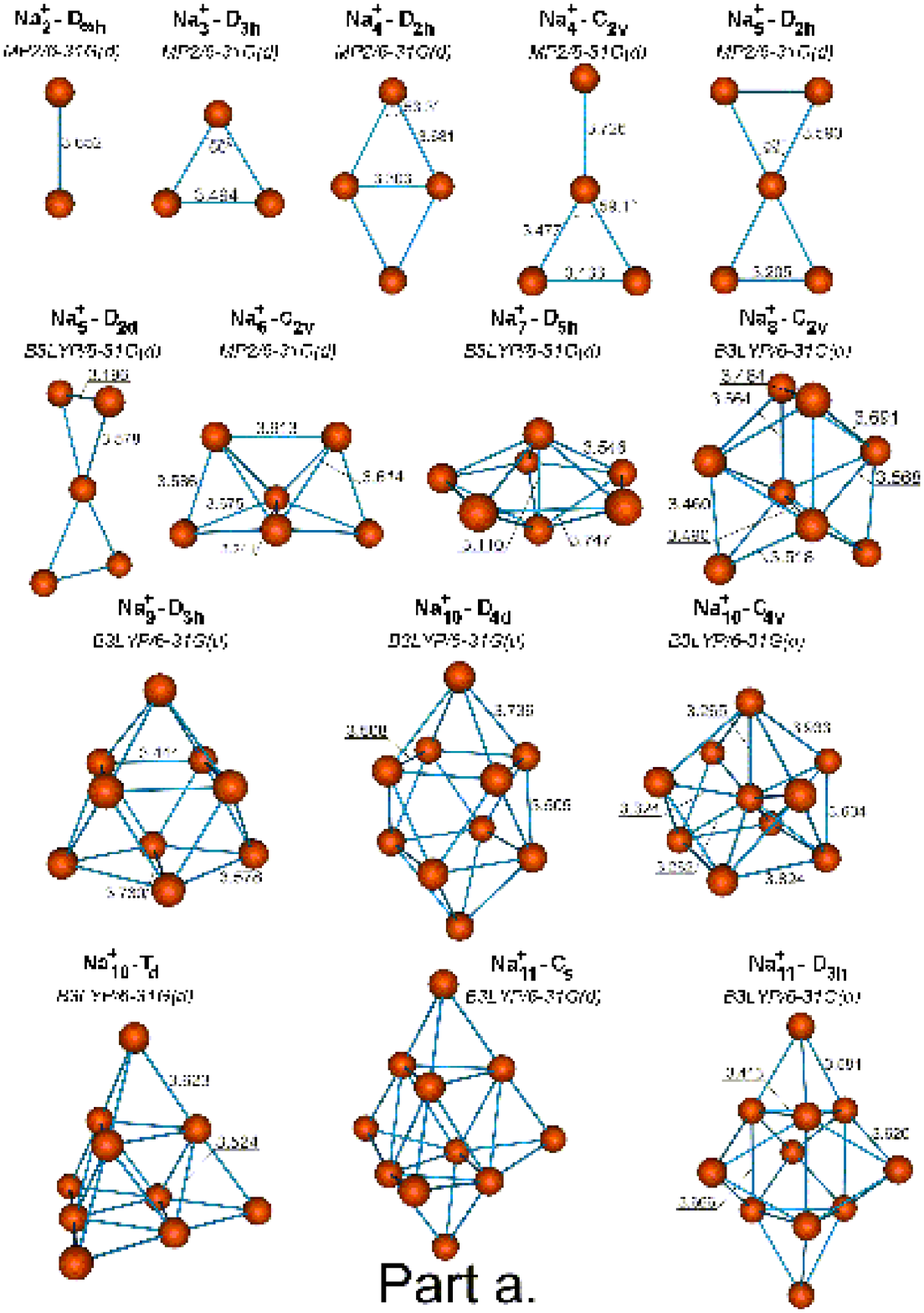}
\end{center}
\end{figure}

\newpage

\newpage

For clusters with $N\leq 6$,
we preferably used the $MP_2$ method.
This method leads to the results, which are in a reasonable agreement 
with those derived by other methods (see e.g. \cite{Bonacic88,Boustani88}).
For example, the side bond length in
the rhomboidal $Na_4$ cluster calculated in \cite{Bonacic88}
by the all-electron Hartree-Fock method is equal to 3.74 \AA,
while in our case it is equal to 3.56 \AA.
The smaller diagonal
value for $Na_4$ is equal to 3.25 \AA\,\, in \cite{Bonacic88},
while we determine it as 3.18 \AA.

The $MP_2$ method becomes more and more computer time demanding
with the growth cluster size. This happens due to
increase in a number of integrals involved in the
computations. It turns out that for larger cluster
systems the $B3LYP$ method is more efficient. The accuracy of
the $B3LYP$ method is comparable with the accuracy of the $MP_2$ method,
as it is clear from the comparison of the $B3LYP$ and $MP_2$
cluster geometries with those
computed in \cite{Bonacic88} by the configuration interaction
method.

Clusters of a certain size can possess various isomer forms, those number
grows dramatically with increasing cluster size.
We illustrate the situation, and calculate several isomers of the
the $Na_3$, $Na_6$, $Na_{10}$, $Na_{11}$ and $Na_{20}$ clusters.
They all are presented in figure \ref{geom_neutral}.
Note, that the linear and equilateral triangular $Na_3$ isomers, have not been
described in the earlier papers \cite{Bonacic88,Boustani87,Boustani88}
(see also \cite{Haberland94,MetCl99,LesHouches}),
in which isosceles triangular isomers  were considered.
The comparison of properties (dipole and quadrupole
moments, total energies, bonding distances) of these clusters will be
given below.

On the example of the $Na_4$ cluster, we demonstrate 
how the  multiplicity of an electronic state of the
system can influence its geometry. Figure \ref{geom_neutral} shows that
the $Na_4$ cluster has the rhomboidal geometry corresponding
to the $D_{2h}$ point symmetry group, if the multiplicity of
the cluster is equal to 1, while, for the multiplicity
being equal to 3, the cluster has the quadratic geometry
characterised by the $D_{4h}$ point symmetry group.

Sodium clusters with $N \leq 5$ have the plane structure, while 
for $N=6$ both plane and spatial isomers are possible. This
feature is consistent with the jellium picture and can be 
explained from the minimization principle for the  cluster surface. Indeed,
the surface of small plane cluster isomers is less in
comparison with the surface of their possible spatial forms.

Comparison of geometries of the neutral and singly-charged
clusters presented in figures \ref{geom_neutral} and \ref{geom_ion}
shows their significant
difference. For smaller sizes ($N \leq 8$), singly-charged
and neutral clusters have sometimes different
point symmetry groups and bonding distances
(see images of the $Na_4$, $Na_5$, $Na_6$ and $Na_8$ clusters and their ions).
The alteration in the geometry of cluster ions occurs due to the
excessive positive charge available in the system.
The structural change of cluster ions becomes less profound
with increasing cluster size, see clusters with $N \geq 10$,
because the excessive positive charge in this case turns out
to be insufficient to produce substantial change in a massive
cluster, although sometimes (compare $Na_{15}$ and $Na_{15}^+$) noticeable
change in the cluster geometry is also possible.

The striking difference in geometries of small singly
charged and neutral
clusters is closely linked to the problem of cluster
fission.  It is natural to assume that with increasing
cluster charge small clusters should become unstable and
fragment into two parts, while for larger cluster sizes
one can expect quasi-stable configurations, which
should decay via the fission process. Calculation of
such configurations is an interesting task, because
it may provide the essential information on the
predominant fission channels in the system. We do not
perform such an analysis in our work, but  
draw attention that geometries of the
cluster ions, like $Na_4^+$,  $Na_5^+$, $Na_6^+$ and $Na_{15}^+$,
lead to the obvious hints on the possible fragmentation
channels in these cluster systems.

Figure \ref{geom_neutral} shows that the clusters $Na_8$ and $Na_{20}$
have the higher
point symmetry group $T_d$ as compared to the other clusters. This
result is in a qualitative agreement with the jellium model.
According to the jellium model \cite{Ekky84,Ekky85,GJ92,IIKZ93,LSSCG00},
clusters with closed shells of delocalized electrons 
have the spherical
shape, while clusters with opened electron shells are
deformed. The jellium model predicts spherical shapes
for the clusters with the magic numbers $N=8, 20, 34, 40 ...$, 
having respectively the following electronic shells filled: 
$ 1s^2 1p^6, 1d^{10}2s^2, 1f^{14}, 2p^6, ...$,.

We have also found the $T_d$ symmetry group isomer
for the $Na_{10}$ cluster. However, this cluster isomer is not
the lowest energy isomer of $Na_{10}$ (see table \ref{table1}). 
The similar situation occurs in the jellium model, where
the $ 1s^2 1p^6, 2s^2$ closed shell electronic configuration
does not minimize the cluster total energy.

Note also, that both the $LDA$ and $HF$ jellium models predict some
deviation from sphericity for  the $Na_{18}$ cluster \cite{LSSCG00}
having $1d$ subshell filled, which is a result of electron
configurations mixing. This fact is also
in a qualitative agreement with the results of our {\it ab initio}
calculations. The point group symmetry of the $Na_{18}$ cluster, $C_{5v}$, is
lower than $T_d$, which is the point symmetry group for the $Na_8$ and 
$Na_{20}$
clusters, and even lower than the point symmetry group for some
opened shell clusters, like $Na_7$ and $Na_{19}$, having the point symmetry
group $D_{5h}$.

Note that there are some clusters possessing relatively low 
point symmetry group, that nevertheless is quite
close to the higher point symmetry group. The higher symmetry
breaking is not occasional and can be explained via the Jahn-Teller effect \cite{LL3}.   
Such situation occurs, for example,
in the $Na_{9}$ and $Na_{11}$ clusters,
which posses the $C_{2v}$ point symmetry group, but their geometry is
close to the geometry of the $D_{3h}$ group.

The jellium prediction on the sphericity of the magic clusters
works not so well for cluster ions. Indeed, the geometry
and the point symmetry group of $Na_9^+$ does not allow one to state
the higher sphericity of this cluster as compared to
its neighbours. The analysis of the quadrupole moments and cluster 
deformations performed below  demonstrates this conclusion quite clearly. This
happens because forces emerging in the cluster during
its transition from neutral to singly charged
state turns out to be insufficient to rearrange the cluster
geometry from  deformed to spherical one.

\begin{figure}[h]
\begin{center}
\includegraphics[scale=0.63]{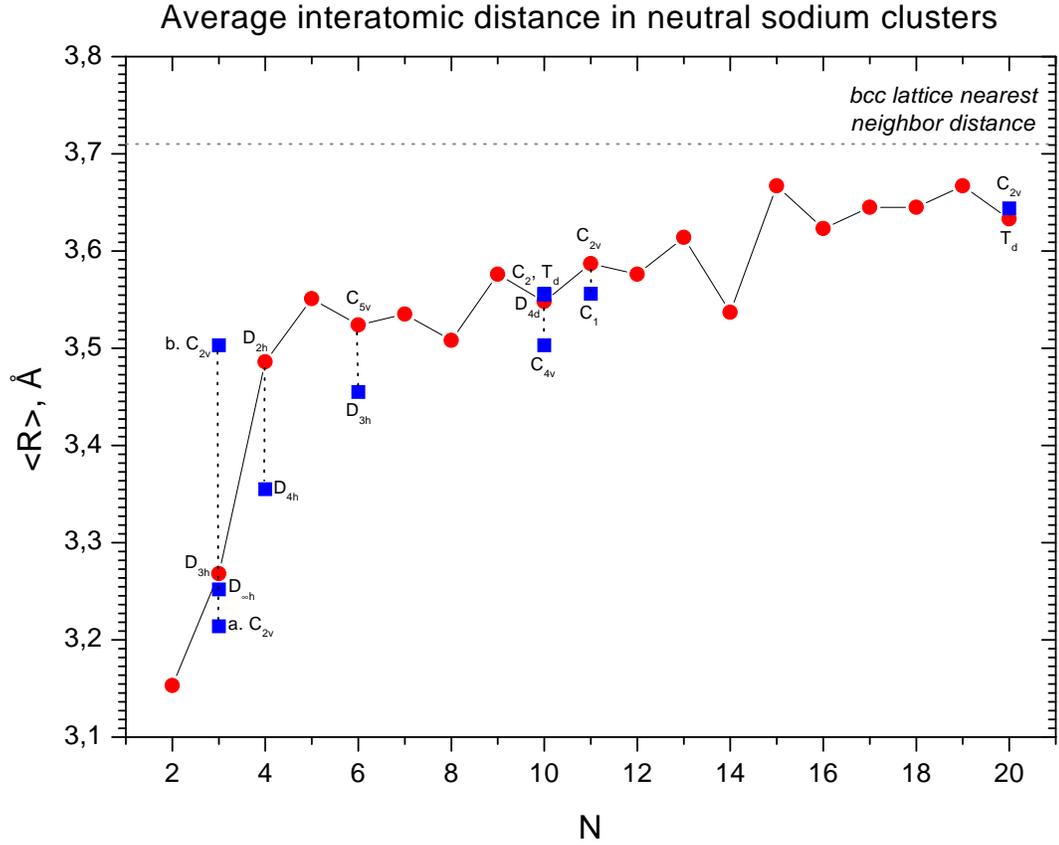}
\end{center}
\caption{Averaged bonding distance  as a function of cluster size
for optimized geometries of neutral sodium clusters.
For some cluster numbers more than one isomer has been
considered. In these cases, labels
indicate  the point symmetry group of the corresponding isomers.
Geometries of the optimized clusters one can find in
figure \ref{geom_neutral}.}
\label{dist_neutral}
\end{figure}

\begin{figure}[h]
\begin{center}
\includegraphics[scale=0.63]{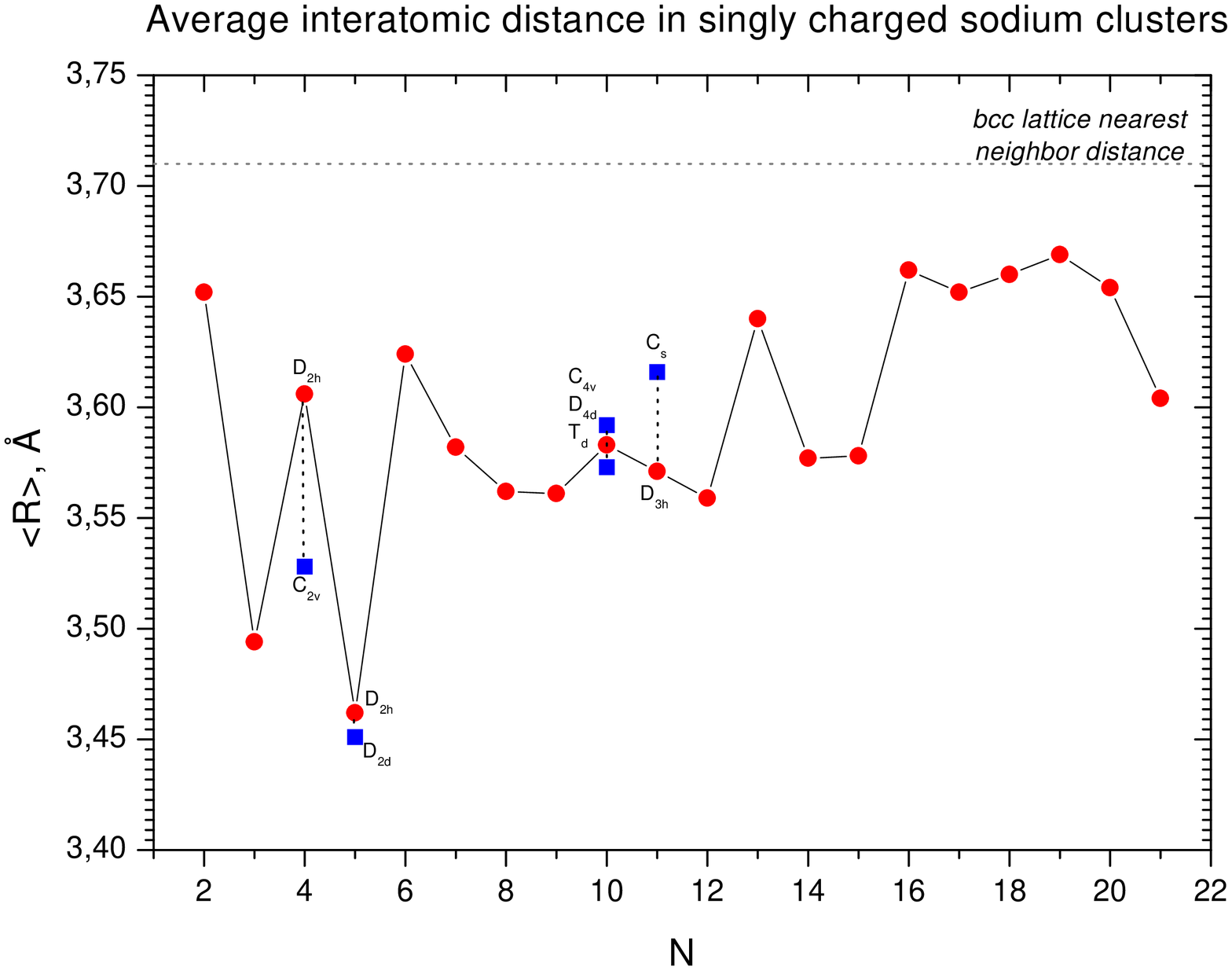}
\end{center}
\caption{Averaged bonding distance  as a function of cluster size
for optimized geometries of singly charged sodium clusters.
For some cluster numbers more than one isomer has been
considered. In these cases, labels
indicate  the point symmetry group of the corresponding isomers.
Geometries of the optimized clusters one can find in
figure \ref{geom_ion}.}
\label{dist_ion}
\end{figure}

We have found two isomers of the $Na_{20}$ cluster, which
have rather regular structure and differ significantly one
from another.  The cluster geometries presented
in figure \ref{geom_neutral} allow one to assume that there exist
at least two independent paths of the cluster structure
formation. Indeed,  the following isomers
$$
Na_6^{C_{5v}} \rightarrow Na_7 
\rightarrow Na_{10}^{C_{4v}} 
\rightarrow Na_{13} \rightarrow
Na_{15} \rightarrow
Na_{16} \rightarrow Na_{17} \rightarrow Na_{18} \rightarrow Na_{19}
\rightarrow Na_{20}^{C_{2v}}
$$
probably belong to the chain
leading  to the formation of the $C_{2v}$  isomer of the $Na_{20}$ cluster,
while the clusters
$$
Na_6^{D_{3h}} \rightarrow Na_8 \rightarrow Na_{9} \rightarrow
Na_{10}^{T_{d}}  \rightarrow
Na_{11}^{C_{1}} \rightarrow Na_{12} \rightarrow Na_{14}  \rightarrow Na_{20}^{T_{d}}
$$
form the path on which the $T_d$ isomer of the $Na_{20}$ cluster
is formed.  Figure \ref{geom_neutral}
clearly shows the steps of the cluster formation process along
these two paths.  Although, for most of N, we have calculated
isomers belonging to one path or another, it is natural
to assume that the two different type of geometries exist
for all N, similar to how it happens for $Na_6$ and
$Na_{20}$ clusters.  For clusters smaller than $Na_6$, one
can not distinguish the two paths clearly enough
as it is seen from figure \ref{geom_neutral}.
Conclusions made  for neutral clusters regarding
the growing process are applicable to
the great extent to singly charged cluster ions
as it is clear from figure \ref{geom_ion}, although
cluster ions geometries sometimes differ substantially
from their neutral prototypes.

Cluster geometries allow one easily to
compute and analyze the average bonding distance as a function
of cluster size. The result of this analysis for
neutral  and singly charged sodium clusters is presented
in figures \ref{dist_neutral} and \ref{dist_ion}.
These figures show how the average bonding distance
converge to the bulk limit indicated in the figures by horizontal lines.
When calculating the average bonding distance in a cluster,
interatomic distances smaller than  4.1 \AA\,\,
have only been considered. This upper limit on the interatomic distances
has been chosen as a distance, which is 10 per cent larger
than to the bcc-lattice nearest neighbour distance in the
bulk sodium.

Figures \ref{dist_neutral} and \ref{dist_ion} show that
the dependence of the average bonding distance,
$\langle R \rangle$, on cluster size
is non-monotonous. For neutral clusters, one can see
odd-even oscillations of $\langle R \rangle$ atop its systematic
growth and approaching the bulk limit. These features
have the quantum origin and  can be explained by the 
delocalization of valence atomic electrons.
Indeed, the odd-even oscillations arise due to the spin paring of
the delocalised electrons. This type of behaviour
is also typical for other cluster characteristics and 
will be discussed below in more detail.
Relatively large increase of the average distance, seen for small
sodium cluster ions with  $N \leq 9$, is also qualitatively
clear. It can be explained by the
Coulomb instability developing in the cluster with increasing its
ionization rate.

\begin{figure}[t]
\begin{center}
\includegraphics[scale=0.63]{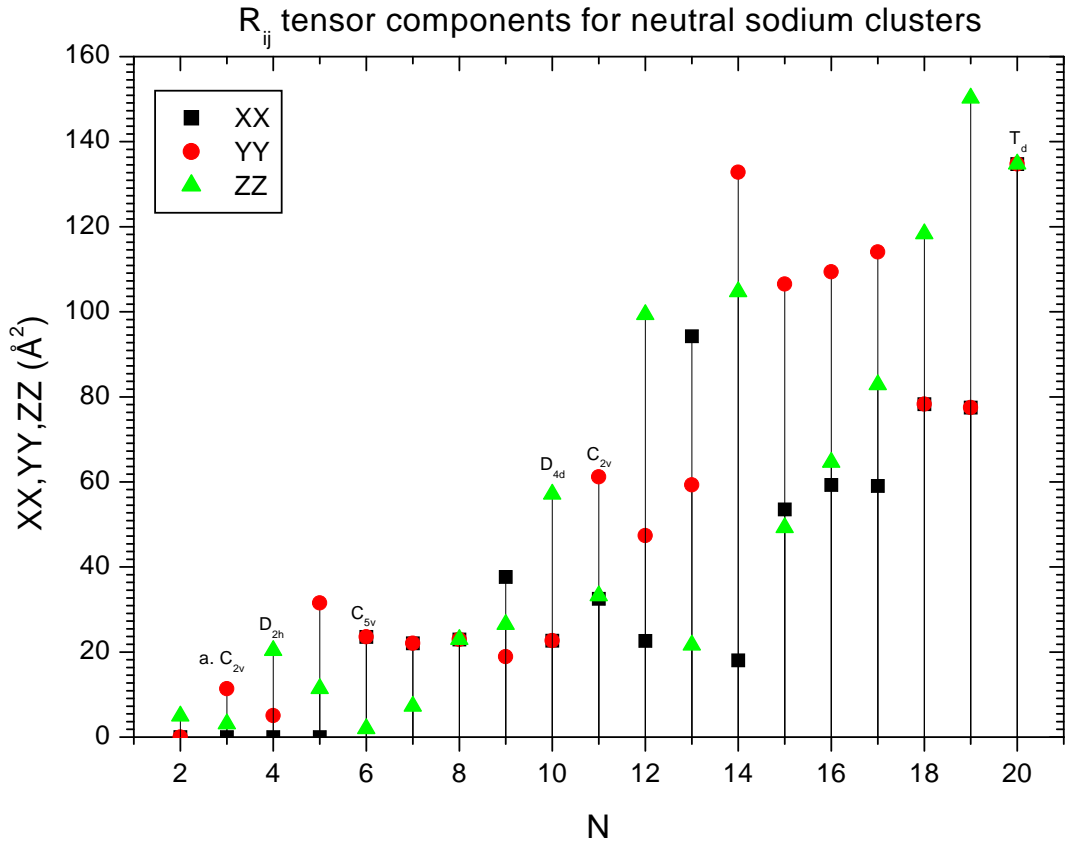}
\end{center}
\caption{The principal values of  tensor $R_{ij}$
for optimized neutral sodium clusters as
a function of cluster size calculated by the
$B3LYP$ method. 
Squares, circles and triangles represent the $R_{xx}$, $R_{yy}$
and $R_{zz}$ tensor principal values respectively.
For some clusters, more than one isomer has been
considered. In these cases, labels
indicate  the point symmetry group of the corresponding isomers.
Geometries of the optimized clusters one can find in
figure \ref{geom_neutral}.}
\label{R_ten_neutral}
\end{figure}

\begin{figure}[h]
\begin{center}
\includegraphics[scale=0.63]{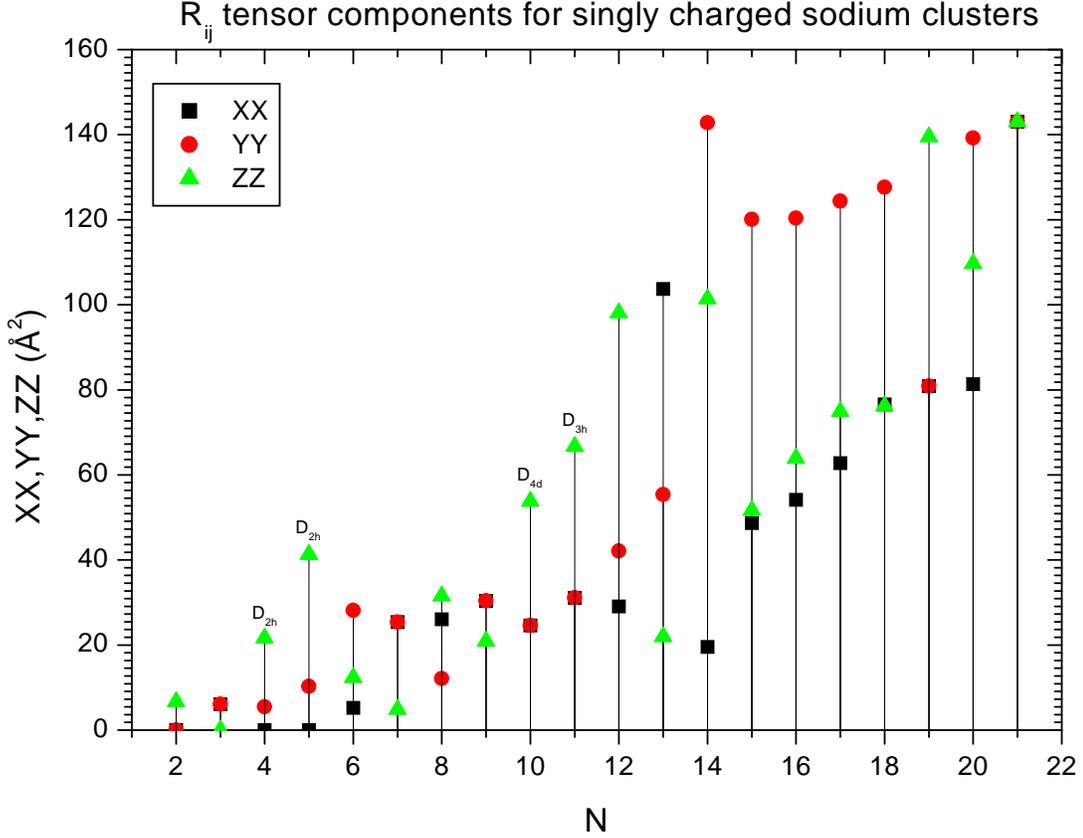}
\end{center}
\caption{The principal values of  tensor $R_{ij}$
for optimized singly charged sodium clusters as
a function of cluster size calculated by the
$B3LYP$ method. 
Squares, circles and triangles represent the $R_{xx}$, $R_{yy}$
and $R_{zz}$ tensor principal values respectively.
For some clusters, more than one isomer has been
considered. In these cases, labels
indicate  the point symmetry group of the corresponding isomers.
Geometries of the optimized clusters one can find in
figure \ref{geom_ion}.}
\label{R_ten_ion}
\end{figure}

Cluster shape can be characterized by the oblate, prolate or
triaxial deformation.
The  prolate deformation of the cluster is characterized
by larger distortion of the ionic charge distribution
along  z-axis as compared
to distortions along  x- and y axes.
In the  oblate deformation case the situation is opposite.
Deformations of the ionic charge distribution
in x- and y- directions are larger than in z-direction.
In both cases the deformations along x- and y- directions
are equal. The triaxial shape deformation is characterized
by unequal distortions of the ionic charge distribution
along  x-, y- and z- directions. Often, however, two of three 
deformations are close  to each other and this allows one
to discuss the triaxially deformed prolate or oblate cases.
Knowledge of the type of the cluster deformation is quite useful for
the comparison with the jellium model results and
the analysis of the metal cluster photon absorption spectra
by metal clusters (see \cite{MetCl99}).

The type of  cluster deformation can be easily determined
by the principle values of the  tensor
$R_{ij}=\sum x_i x_j$.  Here, the summation is performed over
all ions in the system. The principle values of this
tensor $R_{xx}$, $R_{yy}$ and $R_{zz}$ define the dimensions
$R_{x}$, $R_{y}$ and $R_{z}$ 
of the ionic charge distribution in the cluster along the
principle axes  $x$, $y$ and $z$ via the relations:
$R_{x}= \sqrt{R_{xx}/N}$, $R_{y}= \sqrt{R_{yy}/N}$ and
$R_{z}= \sqrt{R_{zz}/N}$. Note that tensor $R_{ij}$ is
closely connected with the cluster moment of inertia tensor
and the quadrupole moment tensor of the ionic distribution. 

In figures \ref{R_ten_neutral} and \ref{R_ten_ion} we
present the principle values  $R_{xx}$, $R_{yy}$ and $R_{zz}$
for a sequence of neutral and singly charged clusters
respectively.
Figures \ref{R_ten_neutral} and \ref{R_ten_ion} demonstrate
how the cluster deformation change as a function of cluster size. 
Figure \ref{R_ten_neutral}
shows that all three principle values are equal
for the tetrahedron group isomers of
the magic clusters $Na_8$ and $Na_{20}$.  
This feature is in the qualitative agreement
with the jellium model, which predicts spherical shapes for the
magic clusters.
In many cases two of three principal values  of $R_{ij}$ are equal or
nearly equal. Using the definition of the prolate and
oblate cluster distortions given above and figures
\ref{R_ten_neutral} and \ref{R_ten_ion}, one can easily
determine the type of cluster deformation. For example,
clusters $Na_{2}$,  $Na_{10}$, $Na_{18}$ and $Na_{19}$
have the prolate deformation along z-principle axis,
because the following condition $R_{xx}= R_{yy} < R_{zz}$
is fulfilled. The clusters
$Na_{6}$ and $Na_{7}$ possess the prolate deformation
because in this case $R_{xx}= R_{yy} > R_{zz}$.
Figures \ref{R_ten_neutral} and \ref{R_ten_ion} show
that most of clusters are triaxially deformed. However,
it is often possible to assign clusters triaxially
deformed prolate or oblate shape, because two of three
principle values are close to each other. Thus, for instance,
$Na_{4}$, $Na_{15}$ are triaxialy prolate clusters, while
$Na_{14}$ is a triaxialy oblate one.
Figures \ref{R_ten_neutral} and \ref{R_ten_ion} also
show the relative value of
prolate and oblate  deformations in various clusters.

One can define a tensor analogous to $R_{ij}$, but for
electrons. We do not plot the principle values of such a tensor
because they are very close in absolute value to the principle values 
shown in figures \ref{R_ten_neutral} and \ref{R_ten_ion} and
could be traced from the principle values of the
cluster total quadrupole moment tensor considered below in subsection 
\ref{q_mom}.

\subsection{Binding energy per atom for small neutral and
singly-charged sodium clusters.}

The binding energy per atom for small neutral and
singly-charged sodium clusters is defined as follows:

\begin{eqnarray}
E_b/N=E_1-E_N/N 
\label{E_b}
\\
E_b^+/N= \left((N-1)E_1+E_1^+-E_N^+\right)/N,
\label{E_b^p}
\end{eqnarray}

\noindent
where $E_N$ and $E_N^+$ are the energies of a neutral  and
singly-charged N-atomic cluster
respectively.
$E_1$ and $E_1^+$ are the energies of a single
sodium atom and an ion.

\begin{figure}[h]
\begin{center}
\includegraphics[scale=0.63]{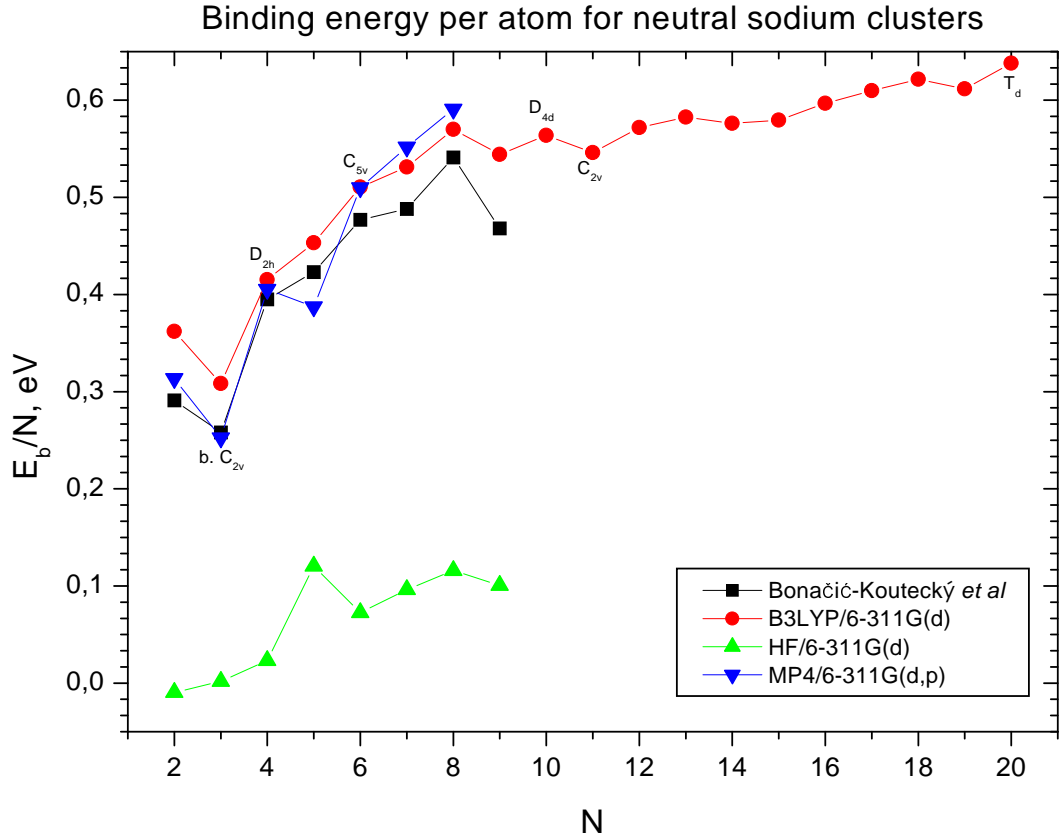}
\end{center}
\caption{Binding energy per atom for neutral sodium clusters as a function of
cluster size.
Circles represent the binding energies per atom calculated
by the $B3LYP$ method,
lower and upper triangles correspond to the 
energies obtained by
the $MP_4$ method and in the
$HF$ approximation respectively. Squares show the result of
the configuration interaction approach from the work
by Bona\v{c}i\'{c}-Koteck\'{y} {\it et al} (for details
see \cite{Bonacic88,Guest88}).
Some points in figure have labels,
indicating the point symmetry group of the isomers represented.
Geometries of the corresponding clusters one can find in
figure \ref{geom_neutral}.}
\label{binding_neutral}
\end{figure}

\begin{figure}[h]
\begin{center}
\includegraphics[scale=0.63]{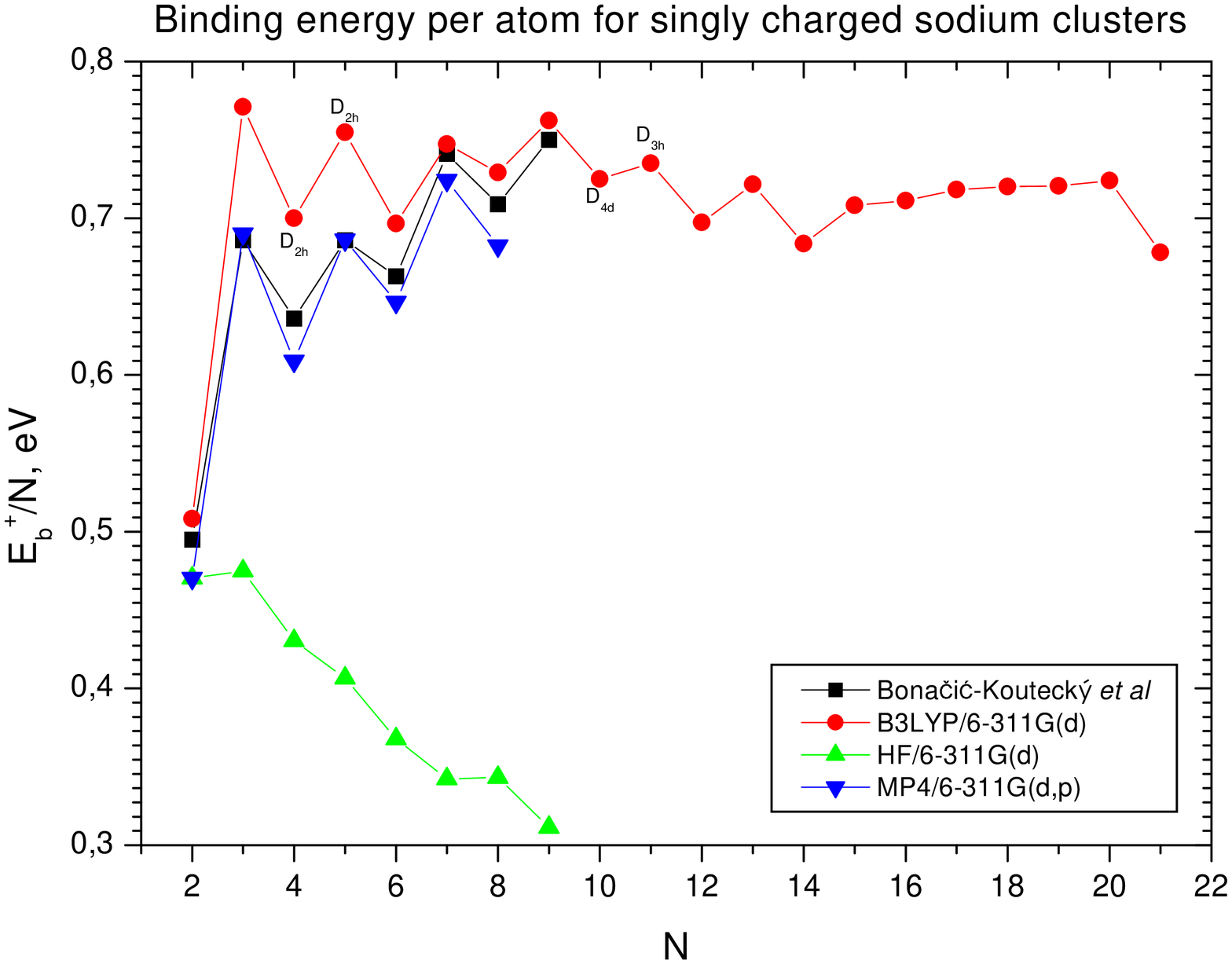}
\end{center}
\caption{Binding energy per atom for singly charged sodium clusters as a function of
cluster size.
Circles represent the binding energies per atom calculated
by the $B3LYP$ method,
lower and upper triangles correspond to the 
energies obtained by
the $MP_4$ method and in the
$HF$ approximation respectively. Squares show the result of
the configuration interaction approach from the work
by Bona\v{c}i\'{c}-Koteck\'{y} {\it et al} (for details
see \cite{Bonacic88,Guest88}).
Some points in figure have labels,
indicating the point symmetry group of the isomers represented.
Geometries of the corresponding clusters one can find in
figure \ref{geom_ion}.}
\label{binding_ion}
\end{figure}

Figures \ref{binding_neutral} and \ref{binding_ion} show the
dependence of the binding energy per atom for neutral and
singly-charged clusters as a function of cluster size.
The energies of clusters have been computed
using the $B3LYP$, $MP_4$ and
$HF$ methods described in section \ref{theory}.
For clusters with $N \leq 8$, computations of the energies have
been performed by the three methods for the sake
of comparison. We wanted to compare the methods by their accuracy
and computation efficiency.
The results of our calculations have also been compared with
those derived  by the configuration interaction ($CI$) method 
in \cite{Bonacic88,Boustani87,Boustani88}).
Figures \ref{binding_neutral} and \ref{binding_ion}
demonstrate that the results of the $MP_4$ and $B3LYP$ methods
are in a reasonable agreement with each other and with the $CI$ results.
The $HF$ points significantly  differ from the $MP_4$, $B3LYP$ and $CI$ ones,
which demonstrates the importance of many-electron correlations, taken
into account in the $MP_4$, $B3LYP$ and $CI$ methods and omitted in the $HF$
approximation.  Note that the energy of $Na_2$,  
if computed in the pure $HF$ approximation, is close to zero,
which means that bonding in this molecule takes place mainly
due to  many-electron correlations.

The energies of clusters larger than $Na_8$ and $Na_8^+$
have been computed by the $B3LYP$ method only, because
this method is more efficient than $MP_4$ and
the accuracy of both methods is comparable.

Figures \ref{binding_neutral} and \ref{binding_ion}
demonstrate the even-odd oscillation behaviour in
the dependence of binding energy on cluster size.
Indeed, for singly charged clusters, odd numbers corresponding to
the singlet multiplicity have higher energies as compared to their even
neighbours. Analogous situation takes place for neutral clusters. In this
case, even cluster numbers have higher binding energies as compared
to their odd neighbours. Note that for neutral clusters this
phenomenon occurs simultaneously with slight systematic growth of the
binding energies per atom with increasing cluster size.

Figures \ref{binding_neutral} and \ref{binding_ion} also
show that the binging energy per atom
in the magic neutral clusters, $Na_8$ and $Na_{20}$, is a little 
higher as compared to other clusters of the close size.
The similar situation takes place for the $Na_9^+$ cluster
in the ionic case. This feature can be qualitatively understood
on the basis of the jellium model: increasing the magic clusters
binding energy takes place due to the 
delocalised electrons shell closure. Note that the binding
energy per atom for the magic $Na_{21}^+$ turns out to be smaller
than that for the neighbouring cluster ions. This happens because
this particular cluster ion isomer is characterized by the $O_h$ point symmetry 
group. Cluster isomers based on this point symmetry group 
usually have the lower binding energy per atom as compared to the
isomers based on the icosahedron point symmetry group like those
with $N \geq 13$ shown
in figures \ref{geom_neutral} and \ref{geom_ion}.

Tables \ref{table1} and \ref{table2}
given in \ref{appendix} provide the accurate values
of the cluster total energies calculated by 
$MP_4$, $B3LYP$ and $HF$ methods.
For neutral clusters with $N \leq 8$, we also present
the cluster energies calculated in \cite{Bonacic88} by the
$CI$ method. The values given in
these tables have been used to plot figures
\ref{binding_neutral} and \ref{binding_ion}.
For some clusters, energies of different symmetry isomers
are also given in the tables.

\subsection{Ionization potentials}

Let us now consider how the ionization potentials of sodium clusters
evolve with increasing  cluster size. Experimentally, such a dependence
has been measured  for sodium clusters in \cite{deHeer93,Akeby90}.

The ionization potential of a cluster consisting of N atoms
is defined as a difference between the energy of the singly-charged
and neutral clusters:

\begin{equation}
IP=E_N^+-E_N 
\label{IP}
\end{equation}

\begin{figure}[h]
\begin{center}
\includegraphics[scale=0.63]{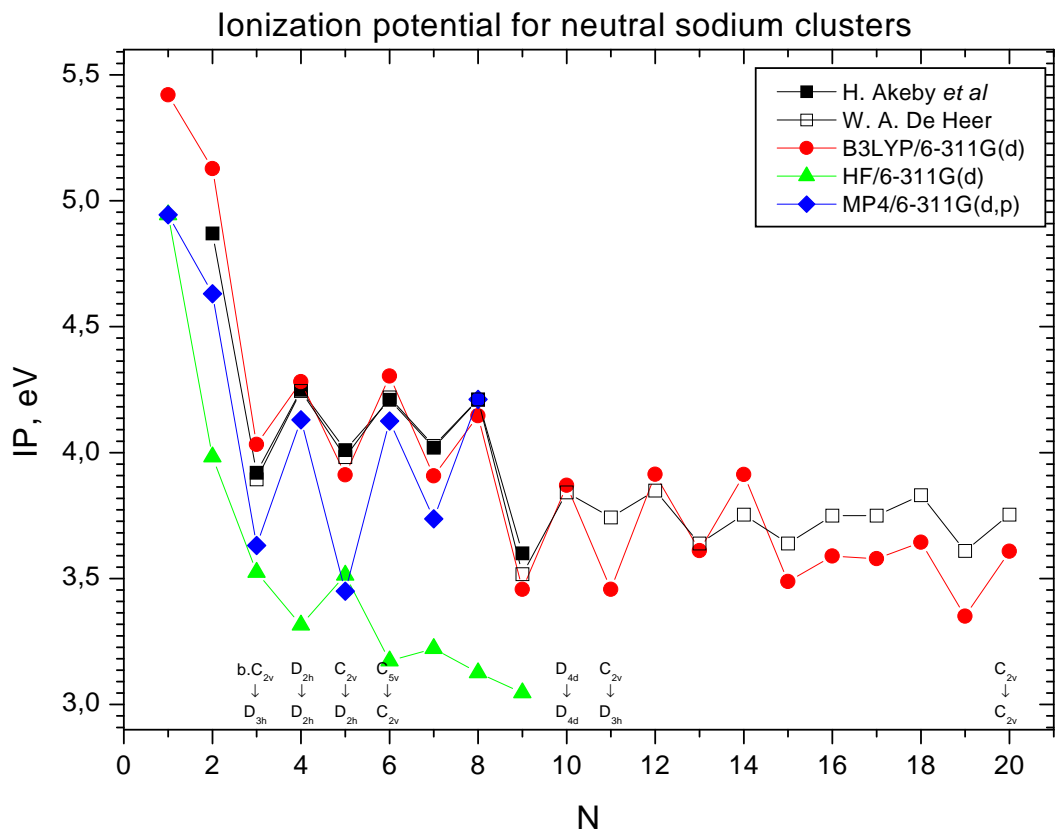}
\end{center}
\caption{Ionization potentials of neutral sodium clusters as a function
of cluster size.
Circles show the results derived by the  $B3LYP$ method.
Triangles and rhomboids represent the ionization potentials
calculated by the $HF$ and $MP_4$ methods respectively. 
Filled and open squares
are the experimental values taken from \cite{Akeby90} and
\cite{deHeer93} respectively. 
For some clusters, more than one neutral and/or 
singly charged cluster isomer has been
considered. In these cases, labels
indicate  the point symmetry group of the initial
neutal and the final charged cluster isomers used
for the calculation of the ionization potential.}
\label{ionization_potential}
\end{figure}

Figure \ref{ionization_potential} shows the dependence of the
clusters ionization potential on N.
Figure \ref{ionization_potential} demonstrates the comparison of the results
derived by different methods, $B3LYP$, $MP_4$ and $HF$ (see section
\ref{theory}), with the  experimental data from  \cite{deHeer93} and
\cite{Akeby90}.
The results of the $B3LYP$ and $MP_4$ methods are in a reasonable
agreement with the experimental data, while the ionization potentials
calculated on the basis of the $HF$ approximation differ substantially
from the experimental observations. 
This comparison shows the role of
many-electron correlations in the formation of
the cluster ionization potentials. The correlation effects are taken into
account by the $B3LYP$ and $MP_4$ methods and omitted in the
$HF$ approximation.

Figure \ref{ionization_potential} demonstrates that the ionisation
potentials drop with increasing cluster size, which is consistent
with predictions of the classical spherical droplet model.
However, this process has many irregularities, which have
quantum origin. Indeed, the dependencies derived by the $MP_4$ and $B3LYP$
methods as well as the experimental one have a prominent
odd-even oscillatory tendency.
The maxima in these dependences correspond to the even-N-clusters, which
means their higher stability  as compared to the neighbouring odd-N-clusters.
This happens because the multiplicities of the even- and odd-N-clusters
are different, being equal to one and two correspondingly.
Interestingly enough that the $B3LYP$ method reproduces correctly
even the experimentally
observed irregularity in the odd-even oscillatory behaviour,
which happens at $N=16$ and $N=17$, and some other minor details
of the experimental data.

A significant step-like decrease in the ionization potential value
happens at the transition from the dimer to the trimer cluster
and also in the transition from $Na_8$ to $Na_9$. Such an irregular
behaviour can be explained by the closure of the electronic
1s- and 1p-shells of the delocalized electrons
in the clusters $Na_2$ and $Na_8$ respectively.
The next significant drop in the ionization
potential value takes place in the transition from the magic
$Na_{20}$ to the $Na_{21}$ cluster.

\subsection{Multipole moments}
\label{q_mom}

\begin{figure}[t]
\begin{center}
\includegraphics[scale=0.63]{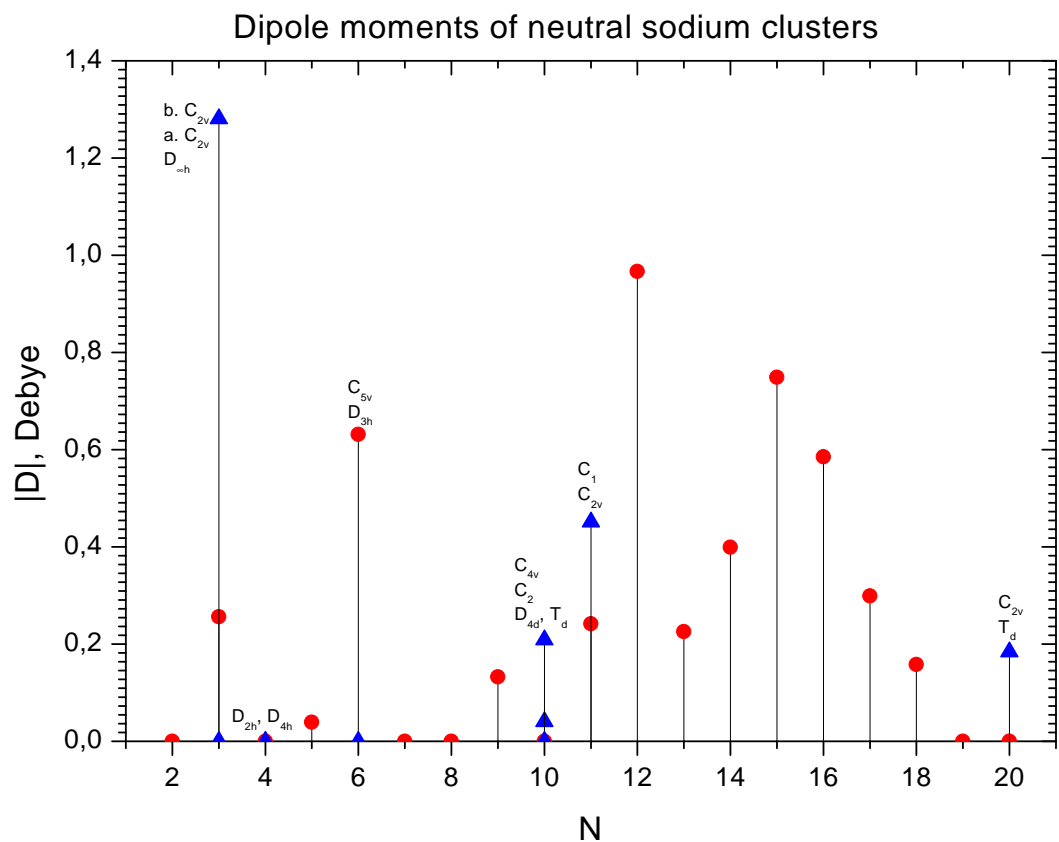}
\end{center}
\caption{Dipole moments of the optimized neutral sodium clusters as
a function of cluster size calculated
by the $B3LYP$  method.
For some clusters, more than one isomer has been
considered. In these cases, labels
indicate  the point symmetry group of corresponding isomers.
Geometries of the optimized clusters one can find in
figure \ref{geom_neutral}.   1 Debye=0.3935 a.u.}
\label{dip_neutral}
\end{figure}

\begin{figure}[h]
\begin{center}
\includegraphics[scale=0.63]{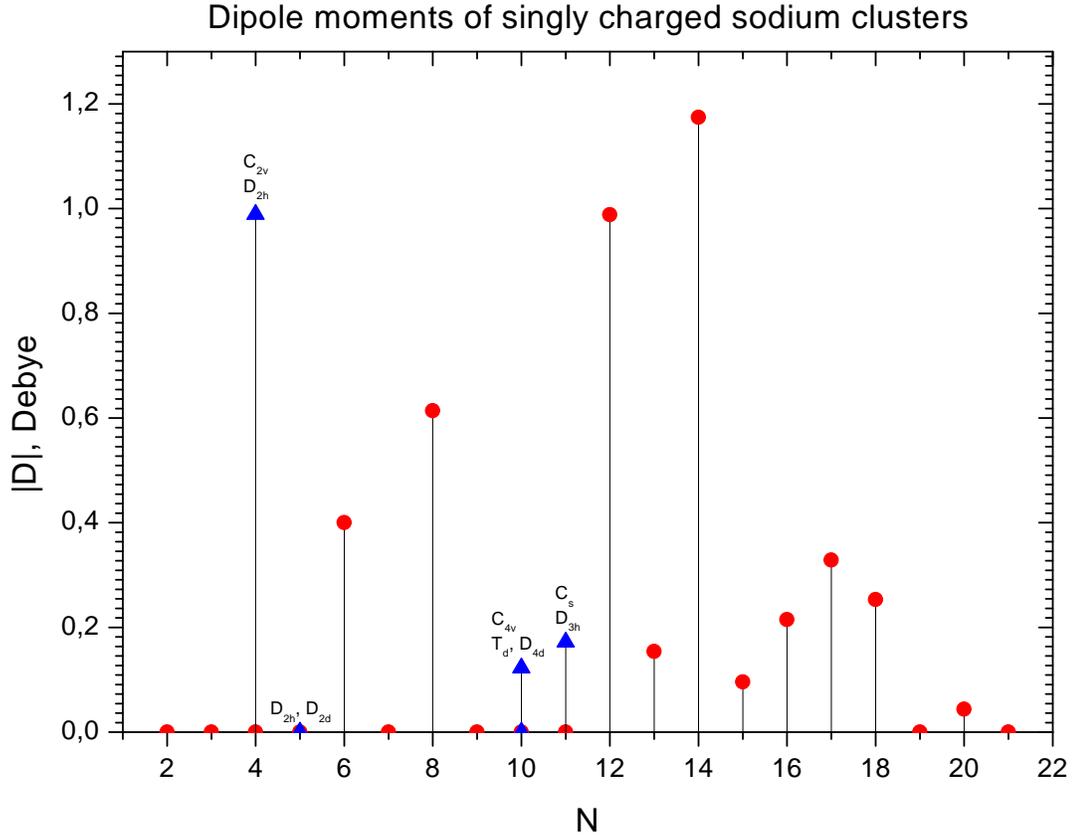}
\end{center}
\caption{Dipole moments for the optimized singly charged sodium clusters as
a function of cluster size calculated by
the $B3LYP$  method.
For some clusters,  more than one isomer has been
considered. In these cases, labels
indicate  the point symmetry group of corresponding isomers.
Geometries of the optimized clusters one can find in
figure \ref{geom_ion}.}
\label{dip_ion}
\end{figure}

We have calculated multipole moments (dipole, quadrupole,
octapole and hexadecapole) for the sodium clusters those
geometry is shown in
figures \ref{geom_neutral} and \ref{geom_ion}. In figures
\ref{dip_neutral} and  \ref{dip_ion}, we plot the absolute values of
the dipole moments
for the neutral and singly charged sodium clusters
as a function of  cluster size.

The dipole moments of some sodium clusters
(see figure \ref{dip_neutral}), which we predict in
our paper, arise due to the fact that the electron charge distribution
not always matches the ionic charge distribution and can be shifted
with respect to the cluster centre of mass. Our calculations show
that only clusters with the C-point symmetry groups, like
the isosceles triangle isomers of $Na_3$, the pentagonal $Na_6$ pyramid
isomer,  $Na_{12}$, $Na_{18}$ and others, possess dipole moments.
These clusters have
either an axis of a certain order or a plane of symmetry, but
no perpendicular symmetry elements (plains or axes). This
rule remains correct even for the $Na_{20}$ cluster isomer with
the symmetry $C_{2v}$, which has
the closed shell configuration $1s^21p^61d^{10}2s^2$ of delocalised
electrons according to the jellium model.
Geometries of the cluster ions differ significantly from the
geometries of the corresponding neutral clusters, but the rule formulated above
on the appearance of the cluster dipole moments remain valid in
this case also as it is clear from figure \ref{dip_ion}.

\begin{figure}[h]
\begin{center}
\includegraphics[scale=0.63]{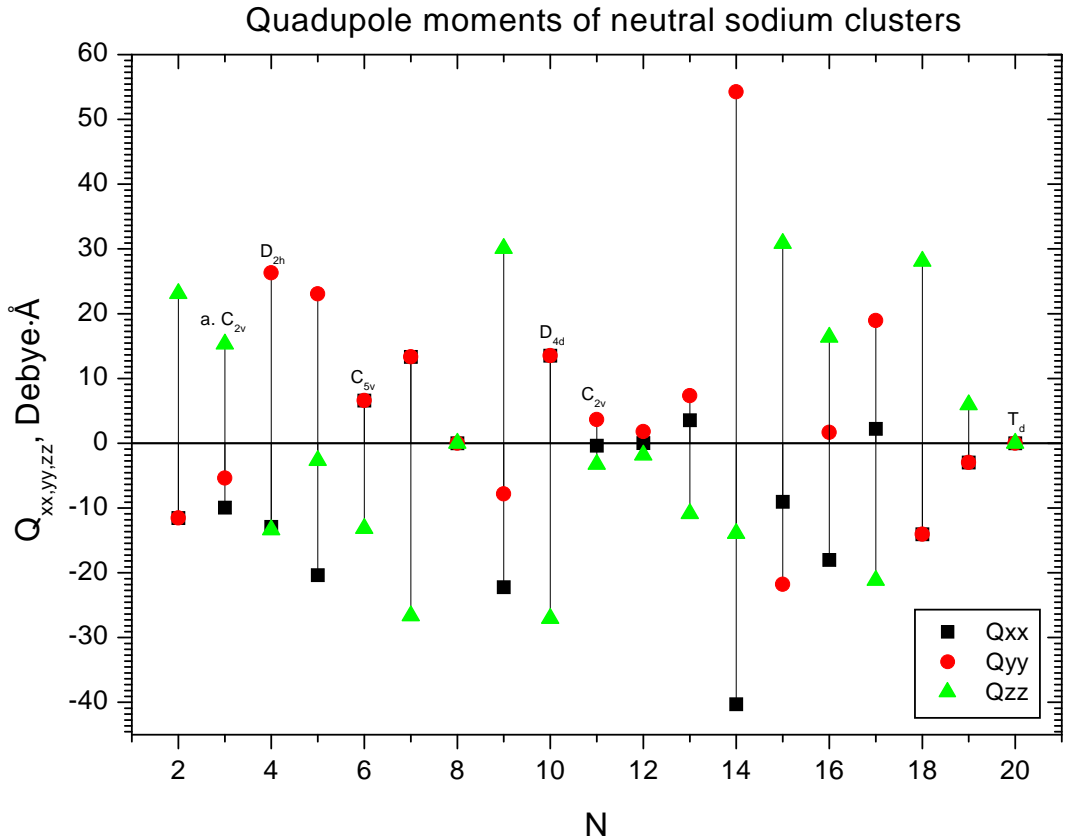}
\end{center}
\caption{The principal values of quadrupole moment tensor
for the optimized neutral sodium clusters as
a function of cluster size calculated by the
$B3LYP$ method. 
Squares, circles and triangles represent the $Q_{xx}$, $Q_{yy}$
and $Q_{zz}$ tensor principal values respectively.
For some clusters, more than one isomer has been
considered. In these cases, labels
indicate  the point symmetry group of corresponding isomers.
Geometries of the optimized clusters one can find in
figure \ref{geom_neutral}.}
\label{quad_neutral}
\end{figure}

\begin{figure}[h]
\begin{center}
\includegraphics[scale=0.63]{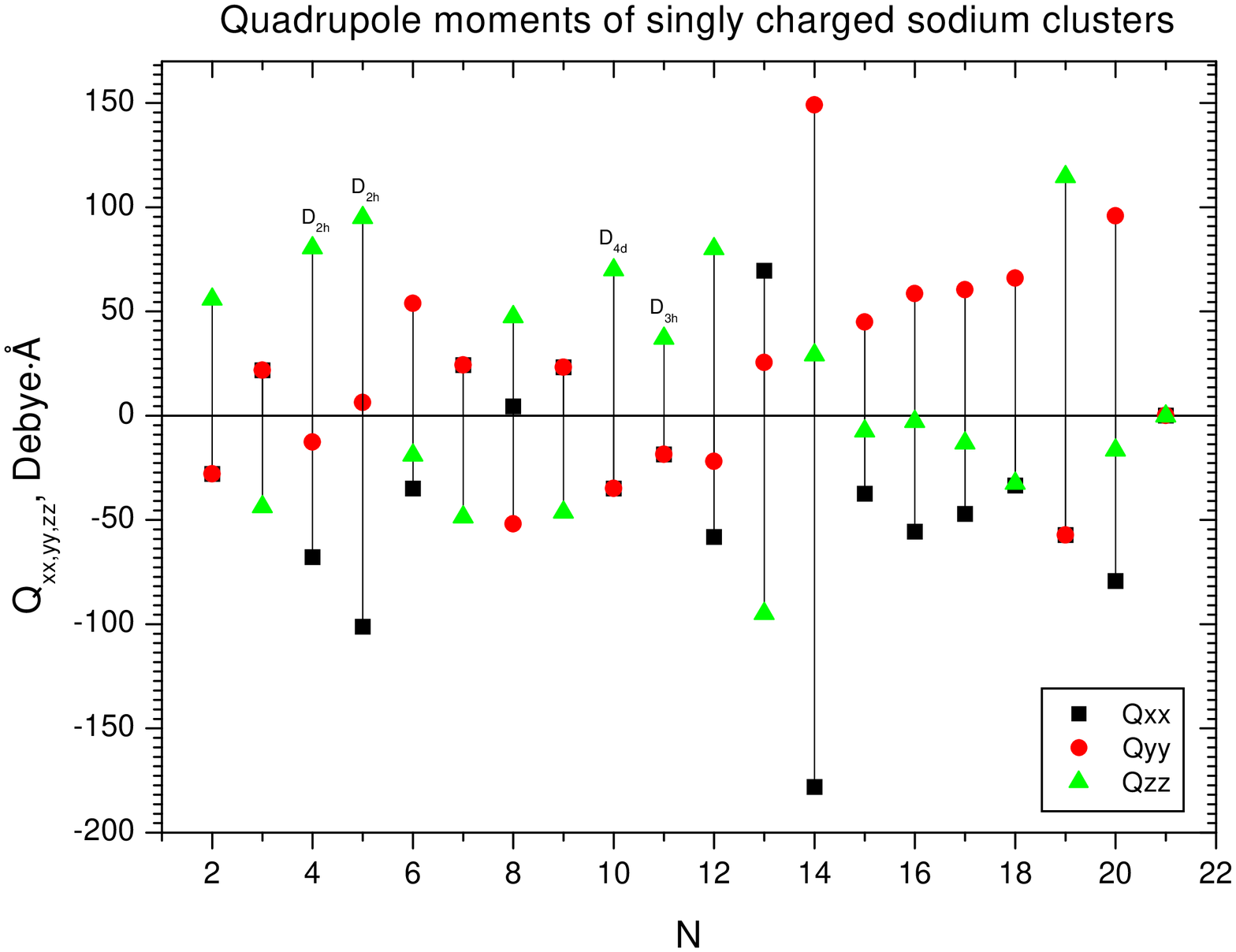}
\end{center}
\caption{The principal values of quadrupole moment tensor
for the optimized singly charged sodium clusters as
a function of cluster size calculated by the
$B3LYP$ method. 
Squares, circles and triangles represent the $Q_{xx}$, $Q_{yy}$
and $Q_{zz}$ tensor principal values respectively.
For some clusters, more than one isomer has been
considered. In these cases, labels
indicate  the point symmetry group of corresponding isomers.
Geometries of the optimized clusters one can find in
figure \ref{geom_ion}.}
\label{quad_ion}
\end{figure}

The principal values of the quadrupole moments tensor 
for the optimized neutral and singly charged clusters
are presented in figures \ref{quad_neutral} and \ref{quad_ion}
respectively.
For clusters with an axis
of symmetry, this axis has been chosen as z-axis of the coordinate
system, in which the calculation of the quadrupole moments has been
performed. 
The quadrupole moment tensor is defined
as an average value of the following operator:
\begin{equation}
Q_{ij}= \sum q (3x_i x_j -\delta_{ij}{\bf r}^2)
\label{quad}
\end{equation}
Here, the summation is performed over all electronic and ionic charges
in the cluster. Note that
the trace of the tensor $Q_{ij}$ is  equal to zero.

The ionic part of $Q_{ij}$ can be expressed via the components
of the tensor $R_{ij}$ discussed in section \ref{geom_opt}.
Note that the knowledge of $Q_{ij}$ and $R_{ij}$ allows
one to construct easily the tensor analogous to $R_{ij}$, but for
electrons. This might be useful for the analysis of  deformations
of electron density distribution in a cluster.

The quadrupole moment tensor can be expressed via the tensor
$\tilde Q_{ij}=\langle\sum q x_i x_j \rangle$,
characterising the averaged dimensions of the total charge distribution.
Here, brackets mean averaging over the electronic charge distribution.
The principal values of the tensor $\tilde Q_{ij}$
should be negative at least for neutral clusters,
because electron density is spilled out of the cluster, which makes
its distribution a little broader than the
distribution of  ions. The similar situation takes place
for cluster ions, but in this case there is non-compensated
positive charge in the system, which brings certain positive
contribution to $\tilde Q_{ij}$ and  makes the principal values
of $\tilde Q_{ij}$ positive in some cases.

The numerical analysis performed in this work shows that 
for neutral sodium clusters
the principal values of $\tilde Q_{ij}$  are always negative, while
for the small cluster ions: $Na_2^+$, $Na_3^+$ and $Na_4^+$ ($C_{2v}$),
some of the principal values are positive.

The principle values of the quadrupole moment tensor
characterize the distortion of the total cluster charge distribution.
Indeed, figure \ref{quad_neutral}
shows that the $Na_8$ and $Na_{20}$ tetrahedron group isomers have the zero
quadrupole moments, which reflect the closeness to sphericity of the
magic clusters.  Our calculations demonstrate that for some open
shell clusters like $Na_{11}$ and $Na_{12}$  the quadrupole moments
turn out to be rather small, although the ionic charge distribution in these
clusters has the prominent deformation as it is clear from
figures \ref{geom_neutral} and  \ref{R_ten_neutral}.
The small quadrupole moments in these clusters is the
result of compensation of the electron and ion components
of $Q_{ij}$.

The quadrupole moments diagram allows one to make 
some conclusions on the type of the shape of the
total charge distribution in a cluster.
The averaged dimensions of the cluster total charge distribution
in x-, y- and z- directions
can be characterized by quantities
$Q^\parallel_z =\tilde Q_{zz} = \langle \sum e z^2 \rangle$,
$Q^\perp_x =  \tilde Q_{xx} =\langle \sum e x^2 \rangle$ and
$Q^\perp_y= \tilde Q_{yy} = \langle\sum e y^2 \rangle$.
Here, the summation is performed over all electrons and ions
in the cluster and brackets mean averaging.
These quantities are connected with the quadrupole
moments tensor defined in (\ref{quad}). 
Indeed, in both the prolate and oblate cases, 
when $Q^\perp_x = Q^\perp_y = Q^\perp$ and $Q^\parallel_z = Q^\parallel$,
the  principal values of the tensor $Q_{ij}$ read as

\begin{eqnarray}
Q_{zz}& = & 2 (Q^\parallel - Q^\perp)
\nonumber
\\
Q_{xx}& = &  ( Q^\perp - Q^\parallel  ) =- \frac{ Q_{zz}}{2}
\nonumber
\\
Q_{yy}&=& Q_{xx} = - \frac{ Q_{zz}}{2}
\label{quad_prol}
\end{eqnarray}

These equations define the important relationships between
the principal values of the quadrupole moments tensor
in the oblate and prolate cases 
and help understanding the quadrupole moments
diagrams shown in figures
\ref{quad_neutral} and \ref{quad_ion}.

Equations (\ref{quad_prol}) show that 
the sign of the principal values  $Q_{xx}$, $Q_{yy}$ and
$Q_{zz}$ depends on the relative value of
$Q^\parallel$ and  $Q^\perp$. With the use of equations (\ref{quad_prol})
and the cluster quadrupole moment diagrams shown in figures \ref{quad_neutral}
and \ref{quad_ion}, one can easily analyse the total charge distribution of
the clusters shown in figures \ref{geom_neutral} and \ref{geom_ion}. Note
that conclusions made on the shape of the total charge distribution and
the shape of ionic component (see figures \ref{R_ten_neutral} and \ref{R_ten_ion})
sometimes differ significantly one from another for some clusters. For example,
the ionic charge distribution in the $Na_{12}$ cluster has a prolate shape, 
while the total charge distribution is oblate.

The quadrupole moments of singly charged sodium clusters
differ substantially from those for the corresponding neutral ones.
The excessive positive charge leads to the rearrangement of
the cluster structure and to the appearance
of the quadrupole moment in the cluster ions like
$Na_8^+$ and $Na_{20}^+$. Although, the electron exchange-correlation
force in a cluster
turns out to be insufficient to change the cluster
geometry  so significantly to make the magic cluster ion $Na_9^+$,
having the closed shell electronic structure of delocalised electron,  
spherical-like without quadrupole moment. Instead,
$Na_9^+$  remains a noticeable deformation.

Let us now discuss the idea for which the cluster multipole moments
play the crucial role and consider the possibility
of the cluster isomers separation by placing the mass selected cluster
beam in the inhomogeneous external field. 
As we have seen from the calculations presented above, different
cluster isomers of the same mass often possess different 
structure and as a result of that different multipole moments (dipole
or quadrupole). However, such cluster isomers are indistinguishable
in the nowadays experiments  with mass selected cluster beams.
They can nevertheless be separated if one puts
the mass selected cluster beam in the inhomogeneous external field.
Let us estimate this effect for the characteristic values
of the dipole and quadrupole moments calculated above.

From the dipole moments diagrams shown in figures \ref{dip_neutral} and
\ref{dip_ion} one can conclude that the
difference in dipole moments for some cluster isomers can be
as large as $1 Debye$ and for the quadrupole often it is about
$40 Debye\cdot$\AA \,or
even larger. The force acting on the cluster with the dipole moment ${\bf D}$
in an external inhomogeneous electric field ${\bf E}({\bf r})$
is equal to \cite{LL2}
\begin{equation}
{\bf F}^D({\bf r})=   {\bf \nabla}
\{  {\bf D} \cdot {\bf E} ({\bf r}) \}.
\label{f_dip}
\end{equation}
\noindent
The components of the force acting on the cluster
with quadrupole moment $Q_{ij}$ is as follows
\cite{LL2}
\begin{equation}
F^Q_{i}({\bf r})=   {\bf \nabla}_i
\{ \frac{  Q_{jk}}{6}  {\bf \nabla}_j  E_k ({\bf r}) \}.
\label{f_quad}
\end{equation}
Here, the summation is assumed over the repeated indices j and k
of the vector and tensor components 
in the right hand side of (\ref{f_quad}).

Let us introduce the  time period $\tau$ during which
the cluster beam passes the inhomogeneous electric field.
One can estimate the distance $\Delta$ on which
isomers will be separated during this period of time
as  $\Delta \sim F \tau^2 /2 M$, where $M$ is the mass 
of the isomer considered and $F$ is the force acting
on either the dipole (see (\ref{f_dip})) or quadrupole
(see (\ref{f_quad})) moment of the cluster.
Substituting
in these equations the characteristic values for the
dipole and  quadrupole moments,
assuming  that the
inhomogeneity of the electric field is about
$\nabla E \sim 5 \cdot 10^3 V/cm^2$, one derives from (\ref{f_dip})
(\ref{f_quad}) that
during the period $\tau \sim 10^{-3} s$ the isomers
with $N=3$ and $\delta D\sim 1 Debye$  become separated
on $\Delta \sim 0.7 mm$ and
that $\Delta \sim  2.8  mm$
for $\delta Q\sim 40 Debye \cdot$\AA, $\tau\sim 10 s$,
$N=5$ and no dipole moment.

These estimates demonstrate that one can create  significant
separation distances for reasonably short periods of time
with the electric field strengths and their gradients achievable
in laboratory conditions.  The experiments with mass
selected and isomer separated  cluster beams
could provide the most accurate information on the structure and
properties of atomic clusters.

\subsection{Polarizabilities}

\begin{figure}[h]
\begin{center}
\includegraphics[scale=0.63]{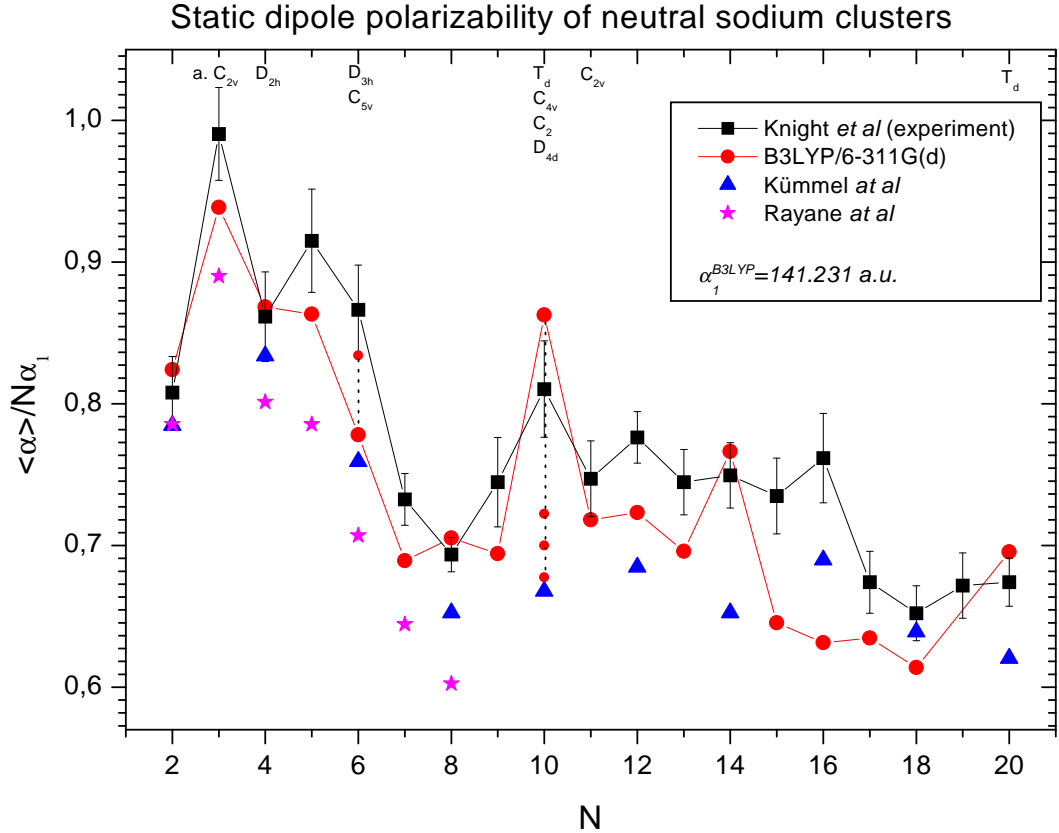}
\end{center}
\caption{Static mean polarizability per atom for neutral sodium clusters normalized
to the polarizability of a single sodium atom.
Circles show the results derived in this work by the $B3LYP$ method.
For some clusters, more than one isomer has been
considered. In these cases, labels
indicate  the point symmetry group of corresponding isomers.
Stars and triangles represent the polarizabilities calculated
in \cite{Rayane99} and \cite{Kummel00} respectively. Squares
are the experimental values taken from \cite{Knight85}.}
\label{polar}
\end{figure}

We have calculated the polarizabilities for the optimized neutral 
sodium clusters (see figure \ref{geom_neutral})  as a function of cluster size.
Results of this calculation are shown in figure  \ref{polar}. 
In this figure, we also plot experimental points from \cite{Knight85}.
Calculation of the polarizabilities has been performed by
the $B3LYP$ method. Figure \ref{polar}
demonstrates quite reasonable agreement of the $B3LYP$ results  with
the experimental data. 

In figure \ref{polar} we also compare the polarizabilities
calculated in our work with those derived by other 
theoretical methods  \cite{Rayane99,Kummel00}.
This figure demonstrates a satisfactory agreement 
of the results of different approaches with each other and 
with the experimental data. This comparison is 
quite important, because in our work  as well as in \cite{Rayane99}
the polarizabilities have been calculated using all electron
{\it ab initio} approach, while in \cite{Kummel00}
they were obtained with the use of pseudopotentials.
Note that our points are closer to the experimental values
than those from \cite{Rayane99}, in spite of the
fact that both calculations have been performed on the basis
of the density functional theory. The difference between the
two schemes of calculation arise in the form of the
density functional and the emploied set of the basis functions.
In \cite{Rayane99}, the so-called  Perdew-Wang-91  density functional 
\cite{PerWan} was used, while we applied its $B3LYP$ form.

Let us also compare the polarizabilities for the $Na_8$ and $Na_{20}$ 
clusters
calculated in the random phase approximation with exchange
in the spherical jellium model,
$\alpha_{Na_8}= 755 a.u.$ and $\alpha_{Na_{20}}= 1808 a.u.$ \cite{Guet95}, 
with our results:
$\alpha_{Na_8}= 797 a.u.$ and $\alpha_{Na_{20}}= 1964 a.u.$ 
The closeness of the values
show that the detailed ionic core structure does not influence
much the value of the clusters polarizabilities. This
comparison shows that the jellium model turns out to be  quite 
a reasonable approximation. 

Figure \ref{polar} shows that  the disagreement
between theoretical and experimental points is not always less than
the experimental error bars. 
Such a disagreement might indicate
that for certain $N$  there have been experimentally detected 
cluster isomers other than those calculated in our work.
For example, the calculated value $\alpha_{Na_6}^{C_{5v}}= 659 a.u.$ 
lies beyond the experimental error bars, while
$\alpha_{Na_6}^{D_{3h}}=706.876 a.u.$ is within the range of the experimental
error.

Note that the polarizabilities of  clusters $Na_{8}$,  $Na_{10}$ 
and $Na_{20}$, possessing the
$T_d$ point symmetry group,
surpass a little the corresponding experimental values, being quite
close to them. For the  $Na_{8}$ and  $Na_{10}$ clusters,
the disagreement of the theoretical and  experimental values 
is within the range of the experimental error.
The similar situation occurs for the  $Na_{14}$ cluster,
characterized by the $C_{2v}$ point symmetry group.
This cluster likely belongs to the cluster chain
leading to the formation of the tetrahedron $Na_{20}$ cluster
from the tetrahedron $Na_{8}$ one 
(see our discussion in section \ref{geom_opt}). Such a situation
allows us to assume that the polarizabilities of
other clusters of this chain, which we
have not analized in this paper, because  they are  
energetically not favorable, will be also 
quite close to the experiment.

\subsection{Normal vibration modes}

\begin{figure}[h]
\begin{center}
\includegraphics[scale=0.63]{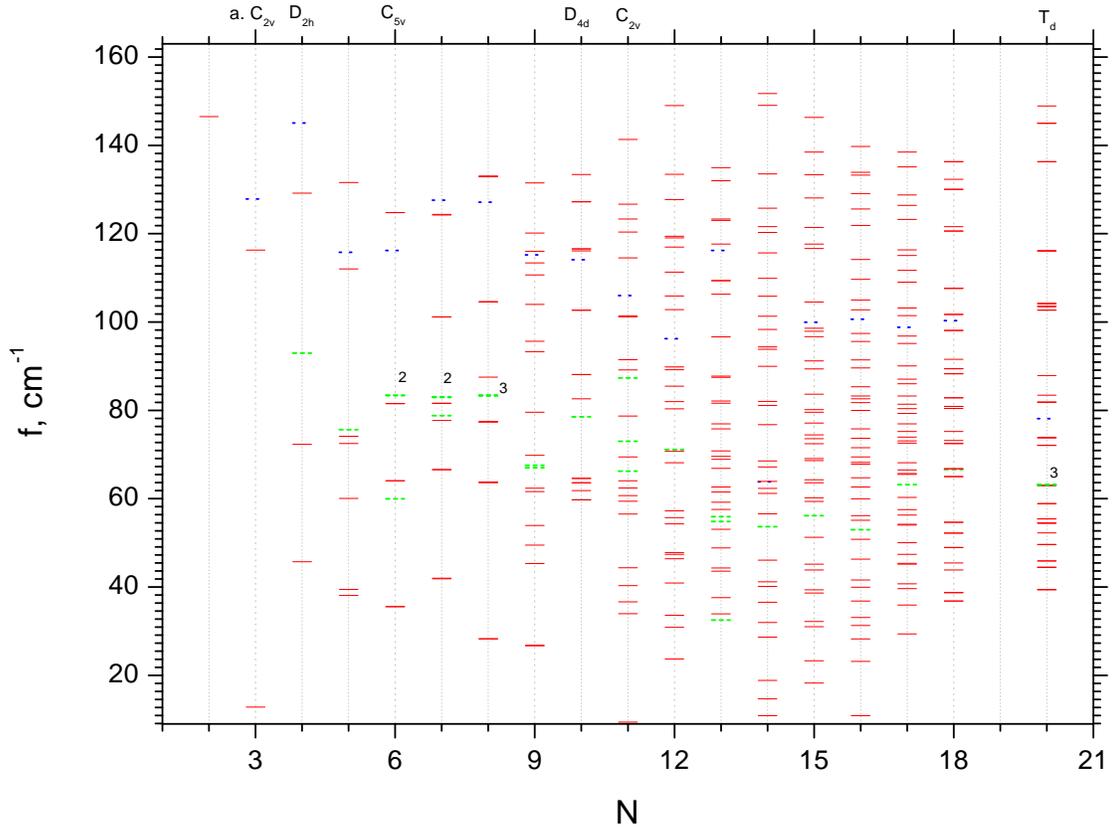}
\end{center}
\caption{Normal vibration frequencies calculated by the $B3LYP$ method for
the neutral sodium clusters with $N \leq 20$.
For each cluster we mark the breathing mode in the spectrum
by dotted line and the surface quadrupole vibration modes
by dashed lines. The number near some of the lines indicate
the degeneracy of the corresponding mode. Note that
we make this only for quadrupole surface vibration modes.}
\label{norm_freq}
\end{figure}

Using the $B3LYP$ method, we have calculated the normal vibration
frequencies for the optimized neutral sodium clusters.
The results of this calculation are shown in figure \ref{norm_freq}. 
In this figure,  we indicate the point symmetry group
for those clusters for which more than one cluster isomer has been
considered (see figure \ref{geom_neutral}).
Numerous
frequencies shown in figure \ref{norm_freq} are degenerate or nearly degenerate.
This explains why the total number of frequencies for most of clusters
is less than the number of vibrational degrees of freedom
available in the system. In the more symmetric clusters, like
$Na_7$, $Na_8$, $Na_{10}$ or $Na_{20}$, 
the rate of generacy of the normal vibration
modes is higher. 

\newpage

\begin{figure}[h]
\begin{center}
\includegraphics[scale=0.45]{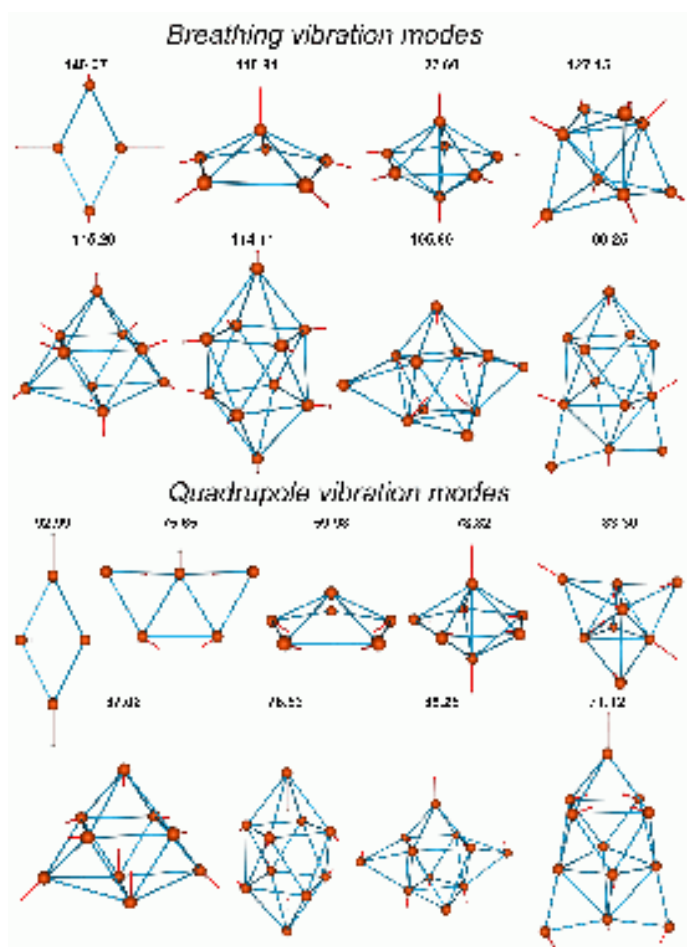}
\end{center}
\caption{Surface and volume vibration modes for the selected
neutral sodium clusters. Number near each cluster image
indicates the frequency of the corresponding normal
vibration mode. The values are given in $cm^{-1}$.}
\label{norm_mod}
\end{figure}

Knowledge of normal vibration modes and their frequencies is important
for physical understanding and quantitative description of the relaxation
of electron plasmon excitations in metal clusters \cite{GISG00}.
One can visualize normal vibration modes, showing the directions
and amplitudes of the atoms displacements by corresponding vectors.
Since it is 
difficult  to show all such pictures in this paper due to their large
number. We focus instead only on the two types of modes breathing and
quadrupole surface vibration modes. Namely these modes have been considered in
\cite{GISG00} within the dynamical jellium model \cite{GSG99} for
the treatment of the electron-phonon coupling in the spherical metal clusters
$Na_{20}$, $Na_{40}$ and $Na_{92}$.

In this paper, we discuss the appearance of these specific vibration
modes in a cluster system and compare their frequencies with the
predictions made in \cite{GISG00} on the basis of the jellium model.
For this purpose, we have analysed all calculated vibration modes
and identified the breathing and three quadrupole vibrations
for each cluster. In figure \ref{norm_mod}, we present images
of the breathing and  quadrupole vibration modes for some clusters
to illustrate the way, how the identification of the modes
has been performed. This figure shows that the identification
made is definite enough.

The results of this analysis are shown in figure \ref{norm_freq},
where for each cluster we mark the breathing mode in the spectrum
by dotted line and the surface quadrupole vibration modes
by dashed lines. The number near some of the lines indicate
the degeneracy of the corresponding modes. Note that
we make this only for quadrupole surface vibration modes.
The degeneracy rate and the number of quadrupole surface
vibration modes  can be easily understood with the help of
the cluster images shown in figure \ref{geom_neutral}.
This figure shows that the prototype of
the breathing mode exists already in the $Na_3$ and  $Na_4$ clusters.
For the $Na_4$ cluster, one can  identify
the quadrupole surface vibration mode, although it is meaningful to
discuss surface vibrations only for the $Na_6$ cluster and larger.
Figure \ref{norm_freq} shows the frequencies of the breathing
and surface vibration modes decrease systematically with
increasing cluster size, although this decrease has
numerous irregularities, particularly for the clusters with
$N < 8$. The frequency of the breathing mode decreases
faster with the growth of $N$ than the frequency of 
the quadrupole surface vibration mode.

Let us compare the calculated frequencies of the breathing and surface
vibration modes  with the predictions
of the jellium model. In  \cite{GISG00}, it was shown
that the breathing vibration mode frequencies
calculated for the
spherical $Na_{20}$, $Na_{40}$ and $Na_{92}$ respectively within the
framework of the dynamical jellium model are quite close
to the values derived from the phonon dispersion law for
metals \cite{Kittel}

\begin{equation}
\Omega^2=\frac{3v_F^2k^2}{M_{Na}(9+k^2v_F^2r_0^3)},
\label{v_freq}
\end{equation}
\noindent
where $M_{Na}=4.2\cdot 10^4$ is the mass of sodium atom, $v_F=(9\pi
/4)^{1/3}/r_0$ is the velocity of cluster electrons on the Fermi
surface, $r_0$ is the Wigner-Seitz radius.
In the long wave limit, equation (\ref{v_freq})
reduces to the Bohm-Staver formula
for the velocity of sound,  $d\Omega /dk=v_F/\sqrt{3M_{Na}}\approx 3\cdot
10^5cm/s$. This number is quite close to the real value
of the velocity of sound in the bulk sodium: $3.2\cdot 10^5cm/s$.

Using the dispersion low (\ref{v_freq}), we estimate the breathing mode
frequencies for  the magic $Na_8$ and $Na_{20}$ clusters. 
The results of this calculation are as follows 
$\Omega_{Na_8}= 104.09 cm^{-1}$, $\Omega_{Na_{20}}= 80.49 cm^{-1}$. 
In this calculation we have used $r_0= 4$.

The frequency values obtained from (\ref{v_freq})
are close to those presented in figure 
\ref{norm_freq}, $\Omega_{Na_8}=127.15 cm^{-1}$, 
$\Omega_{Na_{20}}=78.11 cm^{-1}$.  
The agreement of the frequencies is rather good 
for the  $Na_{20}$ cluster case. 
For $Na_{8}$, the agreement is reasonable, but not
as good as for  $Na_{20}$. Some disagreement arises due
to the fact that the Wigner-Seitz  radius for the $Na_{8}$
cluster is about 10\% smaller than its bulk value. 
Indeed, substituting $r_0=3.6$ in (\ref{v_freq}) one derives
$\Omega_{Na_{8}}=127.10 cm^{-1}$, which is in the nearly
perfect agreement with the {\it ab initio} result.
The decrease of the Wigner-Seitz radius can be easily 
understood from the analysis of the cluster geometry
shown in figure \ref{geom_neutral}.

Now let us compare the quadrupole surface vibration mode
frequencies calculated in our paper (see figure \ref{norm_freq})
with those following from the dynamical jellium model.
According to \cite{GISG00}, the quadrupole surface vibration frequencies,  
$\Omega_2$,
for the spherical $Na_{20}$, $Na_{40}$ and $Na_{92}$ clusters are
equal to
$56.48 cm^{-1}$, $48.41 cm^{-1}$ and $32.28 cm^{-1}$, respectively.
The value of the quadrupole surface vibration frequency
for  the $Na_{20}$ cluster calculated in the present work
is equal to $63.15 cm^{-1}$, which is rather close to the
value predicted in \cite{GISG00}.

The values of the quadrupole surface vibration frequencies  
calculated for  $Na_{20}$, $Na_{40}$ and $Na_{92}$
show relatively slow  decrease 
with the growth cluster size.
Extrapolating these values towards smaller cluster sizes, we
derive frequency values, which are consistent with those
shown in figure \ref{norm_freq}.
This comparison demonstrates that the jellium model calculation of the surface
vibration frequencies  is in a reasonable agreement 
with the more accurate {\it ab initio} many-body theory.
 
The comparison of the jellium model results with those derived by
the more accurate {\it ab initio} many-body theory is important,
because it forms theoretical background for the jellium model
calculations in larger cluster systems, for which {\it ab initio}
methods are hardly possible.
The comparison with the jellium model,
which we performed in this paper, can be extended
towards larger cluster sizes and other collective modes of ions motion.

\section{Conclusion}

In this paper we have calculated the optimized structure and 
various characteristics of sodium clusters consisting of
up to 20 atoms. We have used three different methods: $B3LYP$, $MP_4$
and $HF$. It was demonstrated that the first two methods due to
accounting for  many-electron correlations provide much better
agreement with the available experimental data and theoretical results
based of the configuration interaction method as compared to that
for the Hartree-Fock approximation. This was checked for various
cluster characteristics: cluster geometries, binding energies per atom and
the ionization potentials.

We have also calculated and analyzed the
dependence of the ionic component and total
quadrupole moments of sodium clusters as a function
of their size. It was demonstrated that the cluster shapes
characterized by the quadrupole moments are in a reasonable agreement
with the predictions of the jellium model and the results of
the experimental observations.

We have determined the normal vibration modes and their frequencies
for a number of clusters and demonstrated their qualitative
agreement with the predictions based on the jellium model. 

The results of this work can be extended in various directions.
One can use the similar methods to study structure and
properties of various types of clusters. It is interesting to
extend calculations towards larger cluster sizes and perform
more comparison with the results following from the jellium model
and other simplified theories, based either on pseudopotentials 
or effective interatomic potentials. A lot of novel problems arise,
when considering collisions and electron excitations in the
clusters with the optimized geometries. These and many more 
other problems on atomic cluster physics can be tackled
with the use of methods considered in our work.

\section{Acknowledgements}
The authors acknowledge support from the
INTAS, the Volkswagen Foundation, the Alexander von Humboldt
Foundation and DAAD.


\appendix
\section{Tables}
\label{appendix}

In Appendix, we present tables of the
essential cluster characteristics.  The binding energies per
atom for neutral and singly charged clusters are compiled in
tables \ref{table1} and \ref{table2}.
The principal values of the quadrupole moment tensor for
neutral and singly charged clusters are  presented
in tables \ref{table3} and \ref{table4}.



\begin{table}[h]

\begin{tabular}{|c|c|c|c|c|c|}
\hline
N & \multicolumn{1}{|c}{Symmetry} &   \multicolumn{4}{|c|}{E$_N$ (a.u.)} \\

\hline
     \multicolumn{1}{|c}{ } &
     \multicolumn{1}{|c}{ } &
     \multicolumn{1}{|c}{HF/6-311G(d)} &
     \multicolumn{1}{|c}{MP4/6-311G(d,p)} &
     \multicolumn{1}{|c}{B3LYP/6-311G(d)} &
     \multicolumn{1}{|c|}{Ref: \cite{Bonacic88}} \\

\hline

1 & \multicolumn{1}{|c}{} &
\multicolumn{1}{|c}{-161.8459} &
\multicolumn{1}{|c}{-161.8459} &
\multicolumn{1}{|c}{-162.2866} &
\multicolumn{1}{|c|}{-} \\
\hline

2 & \multicolumn{1}{|c}{$D_{\infty h}$} &
\multicolumn{1}{|c}{-323.6911} &
\multicolumn{1}{|c}{-323.7149} &
\multicolumn{1}{|c}{-324.5999} &
\multicolumn{1}{|c|}{-323.3176} \\
\hline

3 & \multicolumn{1}{|c}{$D_{\infty h}$} &
\multicolumn{1}{|c}{-485.5405} &
\multicolumn{1}{|c}{-485.5626} &
\multicolumn{1}{|c}{-486.8963} &
\multicolumn{1}{|c|}{-} \\
  & \multicolumn{1}{|c}{$_{a.} C_{2v}$} &
\multicolumn{1}{|c}{-485.5403} &
\multicolumn{1}{|c}{-485.5653} &
\multicolumn{1}{|c}{-486.8960} &
\multicolumn{1}{|c|}{-484.9729} \\
  & \multicolumn{1}{|c}{$_{b.} C_{2v}$} &
\multicolumn{1}{|c}{-485.5385} &
\multicolumn{1}{|c}{-485.5656} &
\multicolumn{1}{|c}{-486.8939} &
\multicolumn{1}{|c|}{-} \\
  & \multicolumn{1}{|c}{$D_{3h}$} &
\multicolumn{1}{|c}{-485.5282} &
\multicolumn{1}{|c}{-485.5626} &
\multicolumn{1}{|c}{-486.8889} &
\multicolumn{1}{|c|}{-} \\

\hline

4 & \multicolumn{1}{|c}{$D_{2h}$} &
\multicolumn{1}{|c}{-647.3871} &
\multicolumn{1}{|c}{-647.4433} &
\multicolumn{1}{|c}{-649.2076} &
\multicolumn{1}{|c|}{-646.6494} \\
  & \multicolumn{1}{|c}{$D_{4h}$} &
\multicolumn{1}{|c}{-647.3897} &
\multicolumn{1}{|c}{-} &
\multicolumn{1}{|c}{-649.1965} &
\multicolumn{1}{|c|}{-} \\
\hline

5 & \multicolumn{1}{|c}{$C_{2v}$} &
\multicolumn{1}{|c}{-809.2518} &
\multicolumn{1}{|c}{-809.3008} &
\multicolumn{1}{|c}{-811.5164} &
\multicolumn{1}{|c|}{-808.3174} \\
\hline

6 & \multicolumn{1}{|c}{$C_{5v}$} &
\multicolumn{1}{|c}{-971.0915} &
\multicolumn{1}{|c}{-971.1880} &
\multicolumn{1}{|c}{-973.8324} &
\multicolumn{1}{|c|}{-969.9899} \\
  & \multicolumn{1}{|c}{$D_{3h}$} &
\multicolumn{1}{|c}{-971.0998} &
\multicolumn{1}{|c}{-971.1872} &
\multicolumn{1}{|c}{-973.8344} &
\multicolumn{1}{|c|}{-989.9884} \\

\hline

7 & \multicolumn{1}{|c}{$D_{5h}$} &
\multicolumn{1}{|c}{-1132.9462} &
\multicolumn{1}{|c}{-1133.0634} &
\multicolumn{1}{|c}{-1136.1430} &
\multicolumn{1}{|c|}{-1131.6610} \\
\hline

8 & \multicolumn{1}{|c}{$T_{d}$} &
\multicolumn{1}{|c}{-1294.8015} &
\multicolumn{1}{|c}{-1294.9410} &
\multicolumn{1}{|c}{-1298.4606} &
\multicolumn{1}{|c|}{-1293.3395} \\
\hline

9 & \multicolumn{1}{|c}{$C_{2v}$} &
\multicolumn{1}{|c}{-1456.6466} &
\multicolumn{1}{|c}{-} &
\multicolumn{1}{|c}{-1460.7597} &
\multicolumn{1}{|c|}{-} \\
\hline

10 & \multicolumn{1}{|c}{$C_{2}$} &
\multicolumn{1}{|c}{-} &
\multicolumn{1}{|c}{-} &
\multicolumn{1}{|c}{-1623.0758} &
\multicolumn{1}{|c|}{-} \\
   & \multicolumn{1}{|c}{$D_{4d}$} &
\multicolumn{1}{|c}{-} &
\multicolumn{1}{|c}{-} &
\multicolumn{1}{|c}{-1623.0734} &
\multicolumn{1}{|c|}{-} \\
   & \multicolumn{1}{|c}{$C_{4v}$} &
\multicolumn{1}{|c}{-} &
\multicolumn{1}{|c}{-} &
\multicolumn{1}{|c}{-1623.0554} &
\multicolumn{1}{|c|}{-} \\
   & \multicolumn{1}{|c}{$T_{d}$} &
\multicolumn{1}{|c}{-} &
\multicolumn{1}{|c}{-} &
\multicolumn{1}{|c}{-1623.0530} &
\multicolumn{1}{|c|}{-} \\

\hline

11 & \multicolumn{1}{|c}{$C_{2v}$} &
\multicolumn{1}{|c}{-} &
\multicolumn{1}{|c}{-} &
\multicolumn{1}{|c}{-1785.3737} &
\multicolumn{1}{|c|}{-} \\
   & \multicolumn{1}{|c}{$C_{1}$} &
\multicolumn{1}{|c}{-} &
\multicolumn{1}{|c}{-} &
\multicolumn{1}{|c}{-1785.3726} &
\multicolumn{1}{|c|}{-} \\

\hline

12 & \multicolumn{1}{|c}{$C_{2v}$} &
\multicolumn{1}{|c}{-} &
\multicolumn{1}{|c}{-} &
\multicolumn{1}{|c}{-1947.6917} &
\multicolumn{1}{|c|}{-} \\
\hline

13 & \multicolumn{1}{|c}{$C_{1}$} &
\multicolumn{1}{|c}{-} &
\multicolumn{1}{|c}{-} &
\multicolumn{1}{|c}{-2110.0045} &
\multicolumn{1}{|c|}{-} \\
\hline

14 & \multicolumn{1}{|c}{$C_{2v}$} &
\multicolumn{1}{|c}{-} &
\multicolumn{1}{|c}{-} &
\multicolumn{1}{|c}{-2272.3092} &
\multicolumn{1}{|c|}{-} \\
\hline

15 & \multicolumn{1}{|c}{$C_s$} &
\multicolumn{1}{|c}{-} &
\multicolumn{1}{|c}{-} &
\multicolumn{1}{|c}{-2434.6188} &
\multicolumn{1}{|c|}{-} \\
\hline

16 & \multicolumn{1}{|c}{$C_s$} &
\multicolumn{1}{|c}{-} &
\multicolumn{1}{|c}{-} &
\multicolumn{1}{|c}{-2596.9370} &
\multicolumn{1}{|c|}{-} \\
\hline

17 & \multicolumn{1}{|c}{$C_s$} &
\multicolumn{1}{|c}{-} &
\multicolumn{1}{|c}{-} &
\multicolumn{1}{|c}{-2759.2537} &
\multicolumn{1}{|c|}{-} \\
\hline

18 & \multicolumn{1}{|c}{$C_{5v}$} &
\multicolumn{1}{|c}{-} &
\multicolumn{1}{|c}{-} &
\multicolumn{1}{|c}{-2921.5704} &
\multicolumn{1}{|c|}{-} \\
\hline

19 & \multicolumn{1}{|c}{$D_{5h}$} &
\multicolumn{1}{|c}{-} &
\multicolumn{1}{|c}{-} &
\multicolumn{1}{|c}{-3083.8730} &
\multicolumn{1}{|c|}{-} \\
\hline

20 & \multicolumn{1}{|c}{$T_{d}$} &
\multicolumn{1}{|c}{-} &
\multicolumn{1}{|c}{-} &
\multicolumn{1}{|c}{-3246.2015} &
\multicolumn{1}{|c|}{-} \\
   & \multicolumn{1}{|c}{$C_{2v}$} &
\multicolumn{1}{|c}{-} &
\multicolumn{1}{|c}{-} &
\multicolumn{1}{|c}{-3246.1981} &
\multicolumn{1}{|c|}{-} \\
\hline

\end{tabular}

\caption{
In this  table we present 
the total energies of the optimized neutral sodium clusters.
Numbers of atoms in clusters are given in the first column.
In the second column, the point symmetry groups of clusters
are shown. In the next three columns,
the cluster total energies  derived by the $HF$, $MP_4$
and $B3LYP$ methods are compiled.
For the sake of comparison, the total energies computed 
by the $CI$ method in \cite{Bonacic88}
are presented in
the sixth column.
}
\label{table1}
\end{table}


\newpage

\begin{table}

\begin{center}

\begin{tabular}{|c|c|c|c|c|}
\hline
N & \multicolumn{1}{|c}{Symmetry} &   \multicolumn{3}{|c|}{E$_N^+$ (a.u.)} \\

\hline
     \multicolumn{1}{|c}{ } &
     \multicolumn{1}{|c}{ } &
     \multicolumn{1}{|c}{HF/6-311G(d)} &
     \multicolumn{1}{|c}{MP4/6-311G(d,p)} &
     \multicolumn{1}{|c|}{B3LYP/6-311G(d)} \\

\hline

1 & \multicolumn{1}{|c}{} &
\multicolumn{1}{|c}{-161.6642} &
\multicolumn{1}{|c}{-161.6642} &
\multicolumn{1}{|c|}{-162.0874} \\
\hline

2 & \multicolumn{1}{|c}{$D_{\infty h}$} &
\multicolumn{1}{|c}{-323.5447} &
\multicolumn{1}{|c}{-323.5447} &
\multicolumn{1}{|c|}{-324.4114} \\
\hline

3 & \multicolumn{1}{|c}{$D_{3h}$} &
\multicolumn{1}{|c}{-485.4084} &
\multicolumn{1}{|c}{-485.4322} &
\multicolumn{1}{|c|}{-486.7457} \\
\hline

4 & \multicolumn{1}{|c}{$D_{2h}$} &
\multicolumn{1}{|c}{-647.2653} &
\multicolumn{1}{|c}{-647.2915} &
\multicolumn{1}{|c|}{-649.0502} \\
  & \multicolumn{1}{|c}{C$_{2v}$} &
\multicolumn{1}{|c}{-647.2681} &
\multicolumn{1}{|c}{-647.2919} &
\multicolumn{1}{|c|}{-649.0489} \\
\hline

5 & \multicolumn{1}{|c}{$D_{2h}$} &
\multicolumn{1}{|c}{-809.1226} &
\multicolumn{1}{|c}{-809.1740} &
\multicolumn{1}{|c|}{-811.3727} \\
  & \multicolumn{1}{|c}{$D_{2d}$} &
\multicolumn{1}{|c}{-} &
\multicolumn{1}{|c}{-} &
\multicolumn{1}{|c|}{-811.3629} \\

\hline

6 & \multicolumn{1}{|c}{$C_{2v}$} &
\multicolumn{1}{|c}{-970.9749} &
\multicolumn{1}{|c}{-971.0364} &
\multicolumn{1}{|c|}{-973.6742} \\
\hline

7 & \multicolumn{1}{|c}{$D_{5h}$} &
\multicolumn{1}{|c}{-1132.8278} &
\multicolumn{1}{|c}{-1132.9261} &
\multicolumn{1}{|c|}{-1135.9994} \\
\hline

8 & \multicolumn{1}{|c}{$C_{2v}$} &
\multicolumn{1}{|c}{-1294.6866} &
\multicolumn{1}{|c}{-1294.7863} &
\multicolumn{1}{|c|}{-1298.3082} \\
\hline

9 & \multicolumn{1}{|c}{$D_{3h}$} &
\multicolumn{1}{|c}{-1456.5346} &
\multicolumn{1}{|c}{-} &
\multicolumn{1}{|c|}{-1460.6326} \\
\hline

10 & \multicolumn{1}{|c}{$D_{4d}$} &
\multicolumn{1}{|c}{-} &
\multicolumn{1}{|c}{-} &
\multicolumn{1}{|c|}{-1622.9335} \\
   & \multicolumn{1}{|c}{$C_{4v}$} &
\multicolumn{1}{|c}{-} &
\multicolumn{1}{|c}{-} &
\multicolumn{1}{|c|}{-1622.9278} \\
   & \multicolumn{1}{|c}{$T_{d}$} &
\multicolumn{1}{|c}{-} &
\multicolumn{1}{|c}{-} &
\multicolumn{1}{|c|}{-1622.9273} \\

\hline

11 & \multicolumn{1}{|c}{$D_{3h}$} &
\multicolumn{1}{|c}{-} &
\multicolumn{1}{|c}{-} &
\multicolumn{1}{|c|}{-1785.2509} \\
   & \multicolumn{1}{|c}{$C_{s}$} &
\multicolumn{1}{|c}{-} &
\multicolumn{1}{|c}{-} &
\multicolumn{1}{|c|}{-1785.2455} \\

\hline

12 & \multicolumn{1}{|c}{$C_{2v}$} &
\multicolumn{1}{|c}{-} &
\multicolumn{1}{|c}{-} &
\multicolumn{1}{|c|}{-1947.5479} \\
\hline

13 & \multicolumn{1}{|c}{$C_{1}$} &
\multicolumn{1}{|c}{-} &
\multicolumn{1}{|c}{-} &
\multicolumn{1}{|c|}{-2109.8718} \\
\hline

14 & \multicolumn{1}{|c}{$C_{2v}$} &
\multicolumn{1}{|c}{-} &
\multicolumn{1}{|c}{-} &
\multicolumn{1}{|c|}{-2272.1654} \\
\hline

15 & \multicolumn{1}{|c}{$C_s$} &
\multicolumn{1}{|c}{-} &
\multicolumn{1}{|c}{-} &
\multicolumn{1}{|c|}{-2434.4907} \\
\hline

16 & \multicolumn{1}{|c}{$C_{s}$} &
\multicolumn{1}{|c}{-} &
\multicolumn{1}{|c}{-} &
\multicolumn{1}{|c|}{-2596.8051} \\
\hline

17 & \multicolumn{1}{|c}{$C_{s}$} &
\multicolumn{1}{|c}{-} &
\multicolumn{1}{|c}{-} &
\multicolumn{1}{|c|}{-2759.1222} \\
\hline

18 & \multicolumn{1}{|c}{$C_{s}$} &
\multicolumn{1}{|c}{-} &
\multicolumn{1}{|c}{-} &
\multicolumn{1}{|c|}{-2921.4365} \\
\hline

19 & \multicolumn{1}{|c}{$D_{5h}$} &
\multicolumn{1}{|c}{-} &
\multicolumn{1}{|c}{-} &
\multicolumn{1}{|c|}{-3083.7499} \\
\hline

20 & \multicolumn{1}{|c}{$C_{2v}$} &
\multicolumn{1}{|c}{-} &
\multicolumn{1}{|c}{-} &
\multicolumn{1}{|c|}{-3246.0655} \\
\hline

21 & \multicolumn{1}{|c}{$O_{h}$} &
\multicolumn{1}{|c}{-} &
\multicolumn{1}{|c}{-} &
\multicolumn{1}{|c|}{-3408.3434} \\
\hline

\end{tabular}

\caption{
In this table we present
the total energies of the optimized singly charged sodium clusters.
Numbers of atoms in clusters are given in the first column.
In the second column, the point symmetry groups of clusters
are shown. In the next three columns,
the cluster total energies  derived by the $HF$, $MP_4$
and $B3LYP$ methods are compiled.
}

\label{table2}
\end{center}
\end{table}

\newpage

\begin{table}

\begin{center}

\begin{tabular}{|c|c|c|c|c|}
\hline
     \multicolumn{1}{|c}{N } &
     \multicolumn{1}{|c}{Symmetry} &
     \multicolumn{1}{|c}{$Q_{xx}$, (Debye\AA)} &
     \multicolumn{1}{|c}{$Q_{yy}$, (Debye\AA)} &
     \multicolumn{1}{|c|}{$Q_{zz}$, (Debye\AA)} \\

\hline

2 & \multicolumn{1}{|c}{$D_{\infty h}$} &
\multicolumn{1}{|c}{-11.5622} &
\multicolumn{1}{|c}{-11.5622} &
\multicolumn{1}{|c|}{23.1244} \\
\hline

3 & \multicolumn{1}{|c}{$_{a.} C_{2v}$} &
\multicolumn{1}{|c}{-9.9300} &
\multicolumn{1}{|c}{-5.3883} &
\multicolumn{1}{|c|}{15.3183} \\
  & \multicolumn{1}{|c}{$_{b.} C_{2v}$} &
\multicolumn{1}{|c}{-7.5625} &
\multicolumn{1}{|c}{31.0631} &
\multicolumn{1}{|c|}{-23.5006} \\
  & \multicolumn{1}{|c}{$D_{\infty h}$} &
\multicolumn{1}{|c}{-16.3309} &
\multicolumn{1}{|c}{-16.3309} &
\multicolumn{1}{|c|}{32.6618} \\

\hline
4 & \multicolumn{1}{|c}{$D_{2h}$} &
\multicolumn{1}{|c}{-12.9139} &
\multicolumn{1}{|c}{26.2865} &
\multicolumn{1}{|c|}{-13.3726} \\
  & \multicolumn{1}{|c}{$D_{4h}$} &
\multicolumn{1}{|c}{5.1177} &
\multicolumn{1}{|c}{5.1177} &
\multicolumn{1}{|c|}{-10.2354} \\
\hline

5 & \multicolumn{1}{|c}{$C_{2v}$} &
\multicolumn{1}{|c}{-20.3760} &
\multicolumn{1}{|c}{23.0544} &
\multicolumn{1}{|c|}{-2.6784} \\
\hline

6 & \multicolumn{1}{|c}{$C_{5v}$} &
\multicolumn{1}{|c}{6.5817} &
\multicolumn{1}{|c}{6.5817} &
\multicolumn{1}{|c|}{-13.1634} \\
  & \multicolumn{1}{|c}{$D_{3h}$} &
\multicolumn{1}{|c}{14.4807} &
\multicolumn{1}{|c}{14.4807} &
\multicolumn{1}{|c|}{-28.9614} \\

\hline

7 & \multicolumn{1}{|c}{$D_{5h}$} &
\multicolumn{1}{|c}{13.3285} &
\multicolumn{1}{|c}{13.3285} &
\multicolumn{1}{|c|}{-26.6570} \\
\hline

8 & \multicolumn{1}{|c}{$T_{d}$} &
\multicolumn{1}{|c}{0.0000} &
\multicolumn{1}{|c}{0.0000} &
\multicolumn{1}{|c|}{0.0000} \\
\hline

9 & \multicolumn{1}{|c}{$C_{2v}$} &
\multicolumn{1}{|c}{-22.2202} &
\multicolumn{1}{|c}{-7.8457} &
\multicolumn{1}{|c|}{30.0659} \\
\hline

10 & \multicolumn{1}{|c}{$D_{4d}$} &
\multicolumn{1}{|c}{13.5248} &
\multicolumn{1}{|c}{13.5248} &
\multicolumn{1}{|c|}{-27.0496} \\
   & \multicolumn{1}{|c}{$C_{2}$} &
\multicolumn{1}{|c}{31.6087} &
\multicolumn{1}{|c}{-15.5561} &
\multicolumn{1}{|c|}{-16.0526} \\
   & \multicolumn{1}{|c}{$C_{4v}$} &
\multicolumn{1}{|c}{-14.6949} &
\multicolumn{1}{|c}{-14.6949} &
\multicolumn{1}{|c|}{29.3898} \\
   & \multicolumn{1}{|c}{$T_{d}$} &
\multicolumn{1}{|c}{0.0000} &
\multicolumn{1}{|c}{0.0000} &
\multicolumn{1}{|c|}{0.0000} \\

\hline

11 & \multicolumn{1}{|c}{$C_{2v}$} &
\multicolumn{1}{|c}{-0.3816} &
\multicolumn{1}{|c}{3.6348} &
\multicolumn{1}{|c|}{-3.2532} \\
   & \multicolumn{1}{|c}{$C_{1}$} &
\multicolumn{1}{|c}{-18.4455} &
\multicolumn{1}{|c}{13.9570} &
\multicolumn{1}{|c|}{4.48848} \\

\hline

12 & \multicolumn{1}{|c}{$C_{2v}$} &
\multicolumn{1}{|c}{0.0392} &
\multicolumn{1}{|c}{1.7777} &
\multicolumn{1}{|c|}{-1.8169} \\
\hline

13 & \multicolumn{1}{|c}{$C_{1}$} &
\multicolumn{1}{|c}{3.5616} &
\multicolumn{1}{|c}{7.3169} &
\multicolumn{1}{|c|}{-10.8785} \\
\hline

14 & \multicolumn{1}{|c}{$C_{2v}$} &
\multicolumn{1}{|c}{-40.2978} &
\multicolumn{1}{|c}{54.2376} &
\multicolumn{1}{|c|}{-13.9398} \\
\hline

15 & \multicolumn{1}{|c}{$C_s$} &
\multicolumn{1}{|c}{-9.0476} &
\multicolumn{1}{|c}{-21.7878} &
\multicolumn{1}{|c|}{30.8354} \\
\hline

16 & \multicolumn{1}{|c}{$C_s$} &
\multicolumn{1}{|c}{-18.0272} &
\multicolumn{1}{|c}{1.6718} &
\multicolumn{1}{|c|}{16.3554} \\
\hline

17 & \multicolumn{1}{|c}{$C_s$} &
\multicolumn{1}{|c}{2.2310} &
\multicolumn{1}{|c}{18.9437} &
\multicolumn{1}{|c|}{-21.1747} \\
\hline

18 & \multicolumn{1}{|c}{$C_{5v}$} &
\multicolumn{1}{|c}{-14.0456} &
\multicolumn{1}{|c}{-14.0540} &
\multicolumn{1}{|c|}{28.0996} \\
\hline

19 & \multicolumn{1}{|c}{$D_{5h}$} &
\multicolumn{1}{|c}{-2.9626} &
\multicolumn{1}{|c}{-2.9626} &
\multicolumn{1}{|c|}{5.9252} \\
\hline

20 & \multicolumn{1}{|c}{$T_{d}$} &
\multicolumn{1}{|c}{0.0000} &
\multicolumn{1}{|c}{0.0000} &
\multicolumn{1}{|c|}{0.0000} \\
   & \multicolumn{1}{|c}{$C_{2v}$} &
\multicolumn{1}{|c}{-69.7510} &
\multicolumn{1}{|c}{79.8143} &
\multicolumn{1}{|c|}{-10.0633} \\
\hline

\end{tabular}

\caption{
In this table we present the principal values of the quadrupole
moment tensor calculated for neutral
sodium clusters. The first column shows numbers of atoms in clusters.
The second column gives their point symmetry groups. In the last three
columns, the principal values $Q_{xx}$, $Q_{yy}$ and $Q_{zz}$ are
given. They have been computed by the $B3LYP$ method.
}

\label{table3}
\end{center}
\end{table}

\newpage

\begin{table}
\begin{center}

\begin{tabular}{|c|c|c|c|c|}
\hline
     \multicolumn{1}{|c}{N } &
     \multicolumn{1}{|c}{Symmetry} &
     \multicolumn{1}{|c}{$Q_{xx}$, (Debye\AA)} &
     \multicolumn{1}{|c}{$Q_{yy}$, (Debye\AA)} &
     \multicolumn{1}{|c|}{$Q_{zz}$, (Debye\AA)} \\

\hline

2 & \multicolumn{1}{|c}{$D_{\infty h}$} &
\multicolumn{1}{|c}{-27.9109} &
\multicolumn{1}{|c}{-27.9109} &
\multicolumn{1}{|c|}{55.8218} \\
\hline

3 & \multicolumn{1}{|c}{$D_{3h}$} &
\multicolumn{1}{|c}{21.8547} &
\multicolumn{1}{|c}{21.8547} &
\multicolumn{1}{|c|}{-43.7094} \\
\hline

4 & \multicolumn{1}{|c}{$D_{2h}$} &
\multicolumn{1}{|c}{-67.8170} &
\multicolumn{1}{|c}{-12.6716} &
\multicolumn{1}{|c|}{80.4886} \\
  & \multicolumn{1}{|c}{$C_{2v}$} &
\multicolumn{1}{|c}{-86.4460} &
\multicolumn{1}{|c}{-27.4786} &
\multicolumn{1}{|c|}{113.9246} \\
\hline

5 & \multicolumn{1}{|c}{$D_{2h}$} &
\multicolumn{1}{|c}{-101.2157} &
\multicolumn{1}{|c}{6.2746} &
\multicolumn{1}{|c|}{94.9411} \\
  & \multicolumn{1}{|c}{$D_{2d}$} &
\multicolumn{1}{|c}{-46.6091} &
\multicolumn{1}{|c}{-46.6091} &
\multicolumn{1}{|c|}{93.2182} \\

\hline

6 & \multicolumn{1}{|c}{$C_{2v}$} &
\multicolumn{1}{|c}{-34.9108} &
\multicolumn{1}{|c}{53.8712} &
\multicolumn{1}{|c|}{-18.9604} \\
\hline

7 & \multicolumn{1}{|c}{$D_{5h}$} &
\multicolumn{1}{|c}{24.3267} &
\multicolumn{1}{|c}{24.3267} &
\multicolumn{1}{|c|}{-48.6534} \\
\hline

8 & \multicolumn{1}{|c}{$C_{2v}$} &
\multicolumn{1}{|c}{4.4346} &
\multicolumn{1}{|c}{-51.8751} &
\multicolumn{1}{|c|}{47.4405} \\
\hline

9 & \multicolumn{1}{|c}{$D_{3h}$} &
\multicolumn{1}{|c}{23.1994} &
\multicolumn{1}{|c}{23.1994} &
\multicolumn{1}{|c|}{-46.3988} \\
\hline

10 & \multicolumn{1}{|c}{$D_{4d}$} &
\multicolumn{1}{|c}{-34.9547} &
\multicolumn{1}{|c}{-34.9547} &
\multicolumn{1}{|c|}{69.9094} \\
   & \multicolumn{1}{|c}{$C_{4v}$} &
\multicolumn{1}{|c}{-3.5448} &
\multicolumn{1}{|c}{-3.5448} &
\multicolumn{1}{|c|}{7.0896} \\
   & \multicolumn{1}{|c}{$T_{d}$} &
\multicolumn{1}{|c}{0.0000} &
\multicolumn{1}{|c}{0.0000} &
\multicolumn{1}{|c|}{0.0000} \\

\hline

11 & \multicolumn{1}{|c}{$D_{3h}$} &
\multicolumn{1}{|c}{-18.5476} &
\multicolumn{1}{|c}{-18.5476} &
\multicolumn{1}{|c|}{37.0952} \\
   & \multicolumn{1}{|c}{$C_{s}$} &
\multicolumn{1}{|c}{19.4836} &
\multicolumn{1}{|c}{-10.0197} &
\multicolumn{1}{|c|}{-9.46382} \\
\hline

12 & \multicolumn{1}{|c}{$C_{2v}$} &
\multicolumn{1}{|c}{-58.0823} &
\multicolumn{1}{|c}{-21.8996} &
\multicolumn{1}{|c|}{79.9819} \\
\hline

13 & \multicolumn{1}{|c}{$C_{1}$} &
\multicolumn{1}{|c}{69.5400} &
\multicolumn{1}{|c}{25.4745} &
\multicolumn{1}{|c|}{-95.0145} \\
\hline

14 & \multicolumn{1}{|c}{$C_{2v}$} &
\multicolumn{1}{|c}{-178.1183} &
\multicolumn{1}{|c}{149.0275} &
\multicolumn{1}{|c|}{29.0908} \\
\hline

15 & \multicolumn{1}{|c}{$C_s$} &
\multicolumn{1}{|c}{-37.4527} &
\multicolumn{1}{|c}{44.8752} &
\multicolumn{1}{|c|}{-7.4225} \\
\hline

16 & \multicolumn{1}{|c}{$C_s$} &
\multicolumn{1}{|c}{-55.6664} &
\multicolumn{1}{|c}{58.5058} &
\multicolumn{1}{|c|}{-2.8394} \\
\hline

17 & \multicolumn{1}{|c}{$C_s$} &
\multicolumn{1}{|c}{-47.1267} &
\multicolumn{1}{|c}{60.3728} &
\multicolumn{1}{|c|}{-13.2461} \\
\hline

18 & \multicolumn{1}{|c}{$C_{s}$} &
\multicolumn{1}{|c}{-33.5207} &
\multicolumn{1}{|c}{65.9999} &
\multicolumn{1}{|c|}{-32.4792} \\
\hline

19 & \multicolumn{1}{|c}{$D_{5h}$} &
\multicolumn{1}{|c}{-57.3045} &
\multicolumn{1}{|c}{-57.3045} &
\multicolumn{1}{|c|}{114.6090} \\
\hline

20 & \multicolumn{1}{|c}{$C_{2v}$} &
\multicolumn{1}{|c}{-79.3111} &
\multicolumn{1}{|c}{95.8676} &
\multicolumn{1}{|c|}{-16.5565} \\
\hline

21 & \multicolumn{1}{|c}{$O_{h}$} &
\multicolumn{1}{|c}{0.0967} &
\multicolumn{1}{|c}{0.0967} &
\multicolumn{1}{|c|}{-0.1934} \\
\hline

\end{tabular}

\caption{
In this table we present the principal values of the quadrupole
moment tensor calculated for singly-charged
sodium clusters. The first column shows numbers of atoms in clusters.
The second column gives their point symmetry groups. In the last three
columns, the principal values $Q_{xx}$, $Q_{yy}$ and $Q_{zz}$ are
given. They have been computed by the $B3LYP$ method.
}

\label{table4}
\end{center}
\end{table}


\end{document}